\documentclass[11pt]{article}
\pdfoutput=1
\usepackage[english]{babel}%
\usepackage[utf8]{inputenc}%
\usepackage[T1]{fontenc}%
\usepackage{slashed}
\usepackage[pdftex]{graphicx} %
\usepackage[labelfont={bf}, font=small]{subfig} %
\usepackage{amsmath,amssymb,amsthm,amsfonts,amsbsy,latexsym}
\usepackage{epsf}
\usepackage{hhline} %
\usepackage{cancel}
\usepackage{nicefrac} %
\usepackage{framed}
\usepackage{multirow}
\usepackage[usenames, dvipsnames]{xcolor} %
\usepackage[citecolor=OliveGreen, linkcolor=blue, urlcolor=OliveGreen, colorlinks=true]{hyperref}
\usepackage[font=small,labelfont=bf,format=plain, margin=0em, indention=0em]{caption} %
\usepackage[numbers,sort&compress]{natbib} %
\usepackage{tikz}
\usetikzlibrary{arrows,shapes}%
\usetikzlibrary{trees} %
\usetikzlibrary{decorations.pathmorphing}%
\usetikzlibrary{decorations.markings} %
\usetikzlibrary{calc}%
\usetikzlibrary{decorations.pathreplacing} %
\usetikzlibrary{positioning}
\usetikzlibrary{graphs}
\usetikzlibrary{backgrounds,fit}
\newcommand{\centeron}[2]{{\setbox0=\hbox{#1}\setbox1=\hbox{#2}\ifdim

\wd1>\wd0\kern.5\wd1\kern-.5\wd0\fi \copy0

\kern-.5\wd0\kern-.5\wd1\copy1\ifdim\wd0>\wd1
                                       \kern.5\wd0\kern-.5\wd1\fi}}

\newcommand{\GeV}{\ensuremath{\,\text{GeV}}}
\newcommand{\abs}[1]{\ensuremath{\left|#1\right|}}

\newcommand{\TeV}{\ensuremath{\,\text{TeV}}}
\newcommand{\tn}[1]{\ensuremath{\textnormal{#1}}} 
\newcommand{\ifb}{\text{fb}^{-1}}

\begin{document}
\tikzset{
  x=1pt, y=1pt,
  vector/.style={decorate, decoration={snake}, draw},%
  provector/.style={decorate, decoration={snake,amplitude=2.5pt}, draw},%
  antivector/.style={decorate, decoration={snake,amplitude=-2.5pt}, draw},%
  fermion/.style={draw=black, postaction={decorate}, decoration={markings,mark=at position .55 with {\arrow[draw=black]{>}}}},%
  fermionbar/.style={draw=black, postaction={decorate}, decoration={markings,mark=at position .55
      with {\arrow[draw=black]{<}}}},%
  fermionnoarrow/.style={draw=black},%
  gluon/.style={decorate, draw=black, decoration={coil,amplitude=4pt, segment length=5pt}},%
  scalar/.style={dashed,draw=black, postaction={decorate}, decoration={markings,mark=at position .55
      with {\arrow[draw=black]{>}}}},%
  scalarbar/.style={dashed,draw=black, postaction={decorate}, decoration={markings,mark=at position
      .55 with {\arrow[draw=black]{<}}}},%
  scalarnoarrow/.style={dashed,draw=black},%
  electron/.style={draw=black, postaction={decorate}, decoration={markings,mark=at position .55 with {\arrow[draw=black]{>}}}},%
  bigvector/.style={decorate, decoration={snake,amplitude=4pt}, draw},%
  gluon/.style={decorate, decoration={coil,amplitude=4pt, segment length=5pt}},%
  cbrace/.style={decorate,decoration={brace,amplitude=10pt}, draw},
  action/.style={rectangle, minimum size=11, thick, draw=black, fill=Dandelion},
  decision/.style={rectangle, minimum size=11, thick, dashed, draw=black, fill=Goldenrod, rounded
    corners=4},
  connect/.style={-{stealth}},
 }

\begin{titlepage}
\vspace*{-2.cm}
\begin{flushright}
{\small
DESY 15-252 \\
IPPP/15/77\\
DCPT/15/154
}
\end{flushright}
\vspace*{1.5cm}

\begin{center}
{\Large \bf
Searching for Supersymmetry scalelessly
}
\end{center}
\vskip0.5cm

\renewcommand{\thefootnote}{\fnsymbol{footnote}}

\begin{center} {\large M.~Schlaffer\(^{\,a, b}\), M.~Spannowsky\(^{\,c}\), and A.~Weiler\(^{\, d}\)}
\end{center}

\vskip 20pt

\begin{center}
\centerline{$^{a}${\small \it DESY, Notkestrasse 85, D-22607 Hamburg, Germany}}
\vskip 5pt
\centerline{$^{b}${\small \it Department of Particle Physics and Astrophysics,}}
\centerline{{\small \it  Weizmann Institute of Science, Rehovot 7610001, Israel}}
\vskip 5pt
\centerline{$^{c}${\small \it Institute for Particle Physics Phenomenology, Department of Physics,}}
\centerline{{\small \it Durham University, DH1 3LE, United Kingdom}}
\vskip 5pt
\centerline{$^{d}${\small \it Physik Department T75, James-Franck-Strase 1}}
\centerline{\small\it Technische Universit\"{a}t M\"{u}nchen, 85748 Garching, Germany}

\end{center}

\vglue 1.0truecm

\begin{abstract}
\noindent
In this paper we propose a scale invariant search strategy for hadronic top or bottom plus missing energy final states.
We present a method which shows flat efficiencies and background rejection factors over broad ranges of parameters and
masses. The resulting search can be easily recast into a limit on alternative models. We show the strength of the method
in a natural SUSY setup where stop and sbottom squarks are pair produced and decay into
hadronically decaying top quarks or bottom quarks and higgsinos.
\end{abstract}

\end{titlepage}


\section{Introduction}


Supersymmetric models predict the existence of scalar partners (squarks and sleptons) to the
fermions of the Standard Model. In particular, stops and left-handed sbottom squarks need to be
light in order to solve the Hierarchy problem of the Higgs boson mass. Consequently, Searches for
stops and sbottoms are at the core of the ongoing LHC program. However, despite intense efforts, see
e.g.~\cite{Meade:2006dw,Han:2008gy,Perelstein:2008zt, Plehn:2011tf, Fan:2012jf, Plehn:2012pr,
  Alves:2012ft, Kaplan:2012gd, Han:2012cu, Bhattacherjee:2012ir, Graesser:2012qy,
  Chakraborty:2013moa, Kribs:2013lua, Bai:2013xla, Czakon:2014fka, Ferretti:2015dea,
  Ferretti:2015ala, Fan:2015mxp, Drees:2015aeo, Kobakhidze:2015scd, Kaufman:2015nda, An:2015uwa,
  Hikasa:2015lma, Rolbiecki:2015lsa, Beuria:2015mta, Batell:2015zla, Barducci:2015ffa,
  Crivellin:2015bva, Cohen:2015ala, Eckel:2014wta, Casas:2014eca, Demir:2014jqa, Cohen:2014hxa,
  Katz:2014mba, Mustafayev:2014lqa, Brummer:2014yua, Kim:2014yaa, Brummer:2013upa, Evans:2013jna},
they remain elusive, resulting in stringent limits by dedicated searches performed by both ATLAS
\cite{Aad:2012ywa, Aad:2012xqa, Aad:2012uu, Aad:2014kra, Aad:2014bva, Aad:2014qaa, Aad:2014mfk,
  Aad:2014nra} and CMS \cite{Khachatryan:2014doa, CMS-PAS-SUS-13-009, Chatrchyan:2013xna,
  CMS-PAS-SUS-13-015, CMS-PAS-SUS-14-011, Chatrchyan:2014aea, CMS:2013ida, Chatrchyan:2013fea}.

The interpretation of a new physics search requires a model hypothesis against which a measurement
can be tensioned. Lacking evidence for superpartners and clear guidance from theory, apart from
naturalness considerations, it makes sense to employ search strategies that make as few assumptions
on the model as possible. For most reconstruction strategies a trade-off has to be made between
achieving a good statistical significance in separating signal from backgrounds and the
applicability to large regions in the model's parameter space. Hence, experimental searches are in
general tailored to achieving the best sensitivity possible for a specific particle or decay,
leaving other degrees of freedom of interest unconsidered. This approach can lead to poor
performance for complex models with many physical states and couplings, e.g. the MSSM and its
extensions. Hence, a reconstruction that retains sensitivity over wide regions of the phase space,
thereby allowing to probe large parameter regions of complex UV models, is crucial during the
current and upcoming LHC runs. However, since the number of possible realizations of high-scale
models exceeds the number of analyses available at the LHC, tools like ATOM
\cite{ATOM:20xxaaa,Mahbubani:2012qq,Papucci:2014rja}, CheckMate \cite{Drees:2013wra},
MadAnalysis~\cite{Dumont:2014tja} or FastLim~\cite{Papucci:2014rja} and SModels ~\cite{Kraml:2013mwa} 
have been developed in recent years to recast existing limits on
searches for new physics. A method that shows a flat reconstruction efficiency despite  kinematic
edges and population of exclusive phase space regions would be particularly powerful to set limits
on complex models allowing for broad parameter scans.

\begin{figure}[htb]
  \centering
  \begin{tikzpicture}[baseline=(current bounding box.center), line width=1 pt, scale=40]
    \node at (-1.2,.5) {\(g\)};
    \node at (-1.2,-.5) {\(g\)};
    \node at (2.2,.5) {\(t\)};
    \node at (1.5,-.2) {\(\widetilde{\chi}^-\)};
    \node at (2.5,-.2) {\(\widetilde{\chi}^0\)};
    \node at (1.6,1) {\(\widetilde{\chi}^0\)};
    \node at (1.6,-1) {\(\bar b\)};
    \node at (2.6,-1) {\(\ell^-\)};
    \node at (.5,.7) {\(\tilde{t}\)};
    \node at (.5,-.7) {\(\bar{\tilde{t}}\)};

    \draw[gluon] (-1,0.5)--(0,0.5);%
    \draw[gluon] (-1,-.5)--(0,-.5);%
    \draw[scalarnoarrow] (0,0.5)--(0,-.5);%
    \draw[scalarnoarrow] (0,0.5)--(1,.5);%
    \draw[scalarnoarrow] (0,-0.5)--(1,-.5);%
    \begin{scope}[shift={(1,.5)}]
      \draw[fermionnoarrow] (60:0)--(60:1);%
      \draw[bigvector] (60:0)--(60:1);%
    \end{scope}
    \draw[fermion] (1,.5)--(2,.5);%
    \begin{scope}[shift={(2,-.5)}]
      \draw[fermion] (-60:0)--(-60:1);
    \end{scope}

    \draw[fermionnoarrow] (1,-.5)--(2,-.5);
    \draw[bigvector] (2,-.5)--(1,-.5);
    \draw[fermionnoarrow] (2,-.5)--(3,-.5);
    \draw[bigvector] (2,-.5)--(3,-.5);
    \begin{scope}[shift={(1,-.5)}]
      \draw[fermionbar] (-60:0)--(-60:1);
    \end{scope}
  \end{tikzpicture}
  \caption{Generic Feynman diagram for stop production and decay. The initially produced squarks in
    our setup can be any of the two stops or the lighter sbottom. They decay subsequently into a top
    or a bottom and a higgsino, such that the electric charge is conserved.}
  \label{fig:Feynman_diag}
\end{figure}
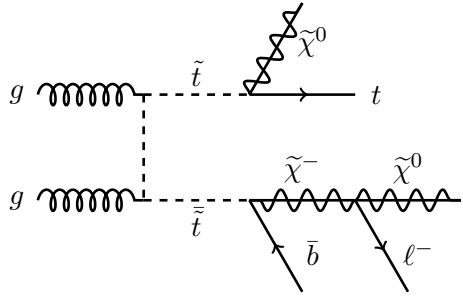

In this paper we develop a reconstruction strategy for third-generation squarks that accumulates sensitivity from a wide range of {\it different phase space regions} and for a {\it variety of signal processes}\footnote{We significantly expand on the proposals of \cite{Gouzevitch:2013qca} and \cite{Azatov:2013hya}. In \cite{Gouzevitch:2013qca} a flat reconstruction efficiency was achieved over a wide range of the phase space for, however, only one process, $pp \to HH \to 4b$, and only a one-step resonance decay, i.e. $H\to \bar{b}b$. The authors of \cite{Azatov:2013hya} showed that in scenarios with fermionic top partners fairly complex decay chains can be reconstructed including boosted and unboosted top quarks and electroweak gauge bosons. While several production modes were studied that can lead to the same final state, only one final state configuration was reconstructed. Focusing on a supersymmetric cascade decay, we will show that a flat reconstruction efficiency can be achieved over a wider region of the phase space, for a variety of production mechanisms and final state configurations.}. The proposed reconstruction is therefore a first step towards an general interpretation of data, i.e.~recasting. As a proof of concept, we study stop and sbottom pair production, followed
by a direct decay into a hadronically decaying top or a bottom and a neutralino or chargino, see Fig.~\ref{fig:Feynman_diag}. In our analysis we use simplified topologies, including only sbottom and stops as intermediate SUSY particles and we focus on jets and missing energy as final state signal. While this example might be oversimplifying, e.g. it might not capture long decay chains that arise if the mass spectrum is more elaborate, this setup is motivated by naturalness \cite{Barbieri:1987fn,deCarlos:1993rbr,Anderson:1994dz}, i.e.~it resembles minimal, natural spectra with light stops, higgsinos and gluinos where all other SUSY
particles are decoupled \cite{Papucci:2011wy}.\par
To be more specific, within this simplified setup the shape of the event depends strongly on the mass difference between the
initially produced squarks and the nearly mass degenerate higgsinos in comparison to the top quark
mass $\mathcal{Q}=(m_{\tilde q}-m_{\tilde h})/m_t$. We can identify three regions in the
physical parameter space leading to distinctly different topologies.
\begin{enumerate}
\item $\mathcal{Q}<1$: The only accessible two-body decay of
  the produced squark is the decay into a bottom quark and a charged higgsino. Possible
  three-body decays contribute only in small areas of the parameter space.
\item $\mathcal{Q}\gtrsim 1$: The decay into a top quark and a higgsino can become the main decay
  channel for the produced squark, depending on the squark and the parameter point. The top quark of
  this decay will get none or only a small boost from the decay and thus its decay products might
  not be captured by a single fat jet. However, the intermediate $W$ boson can lead to a two-prong
  fat jet that can be identified by the BDRS tagger \cite{Butterworth:2008iy}.
\item $\mathcal{Q}\gg 1$: If the squark decays into a top quark the latter will be very boosted
  and its decay products can no longer be resolved by ordinary jets. Yet they can be captured by one
  fat jet and subsequently identified as a decaying top by the HEPTopTagger
  \cite{Plehn:2009rk,Plehn:2010st}. The HEPTopTagger was designed to reconstruct mildly to highly boosted top quarks in final states with many jets, as anticipated in the processes at hand. However, other taggers with good reconstruction efficiencies and low fake rates in the kinematic region of $\mathcal{Q} \gg 1$ (see e.g. \cite{Abdesselam:2010pt, Plehn:2011tg, Altheimer:2012mn} and references therein) can give similar results.
\end{enumerate}

Because the value of $\mathcal{Q}$ is unknown and the event topology crucially depends on it, a
generic reconstruction algorithm that is insensitive to details of the model needs to be scale
invariant, i.e.~independent of $\mathcal{Q}$. Hence, it needs to be able to reconstruct individual
particles from the unboosted to the very boosted regime.

Apart from scanning a large region of the parameter space, such an analysis has the
advantage that it captures the final state particles from the three possible intermediate squark states $\tilde t_{1,2}$ and $\tilde b_1$, even if they have different masses. Therefore the effective signal
cross section is increased compared to a search strategy which is only sensitive to specific processes and within a narrow
mass range. In order to preserve this advantage we furthermore apply only cuts on variables that are
independent of $\mathcal{Q}$ or the mass of one of the involved particles.\par

The remainder of this paper is organized as follows. In Section \ref{sec:event-generation} we give
details on the parameter space that we target and the signal and background event
generation. Section \ref{sec:analysis-1} contains a thorough description of the reconstruction of
the top quark candidates and the proposed cuts as well as the results of the analysis. We conclude
in Section \ref{sec:final-remarks-1}.

\section{Event generation}
\label{sec:event-generation}

\subsection{Signal sample and parameter space}
\label{sec:signal-sample}

As explained in the previous section, not only $\tilde t_1$ but also $\tilde t_2$ and $\tilde b_1$
production contributes to the signal. For all these three production channels we consider the decay
into a higgsino $\widetilde \chi_1^\pm$, $\widetilde \chi_{1,2}^0$ and a top or bottom quark. Since
in our simplified topology setup we assume the higgsinos to be mass degenerate, we generate only the
decay in the lightest neutralino $\tilde q\to q+\widetilde\chi_1^0$ and the chargino
$\tilde q \to q' + \widetilde \chi_1^\pm$, where $q$, $q'$ stand for $t$ or $b$. The decay of the
second lightest neutralino $\widetilde \chi_2^0\to \widetilde \chi_1^{0,\pm} + X$ and of the
chargino to one of the neutralinos $\widetilde \chi_1^\pm \to \widetilde \chi_{1,2}^0+X$ does not
leave any trace in the detector since the emitted particles $X$ will be extremely soft. Thus, the
event topologies for $\tilde t_{1} \to t+\widetilde \chi_{1,2}^{0}$ will be the same and the
different cross section for this topology can be obtained by rescaling with appropriate
branching ratios.
\par
We consider the following points in the MSSM parameter space. At fixed $A_t=200\,\GeV$ and
$\tan \beta=10$ we scan in steps of $50\,\GeV$ over a grid defined by
$\mu \leq m_{Q_3}, m_{u_3} \leq 1\,400\,\GeV$ for the two values of $\mu=150, 300\,\GeV$.  The
gaugino masses as well as the other squark mass parameters are set to $5\,\TeV$ while the remaining
trilinear couplings are set to zero. For each grid point we calculate the spectrum and the branching
ratios with \texttt{SUSY-HIT} \cite{Djouadi:2006bz}. Despite the specific choices for the parameters
our results will be very generic. An increased $A_t$ would enhance the mixing between the left- and
right-handed stops and thus render the branching ratios of the physical states into top and bottom
quarks more equal. However, since the reconstruction efficiencies for both decay channels are
similar, the final results would hardly change. The change due to a different mass of the physical
states can be estimated from our final results. Similarly, a different choice for $\mu$ only shifts
the allowed region in the parameter space and the area where the decay into a top quark opens up but
does not affect the efficiencies.

\begin{figure}[tbh!]
  \centering
  \includegraphics[width=0.44\textwidth]{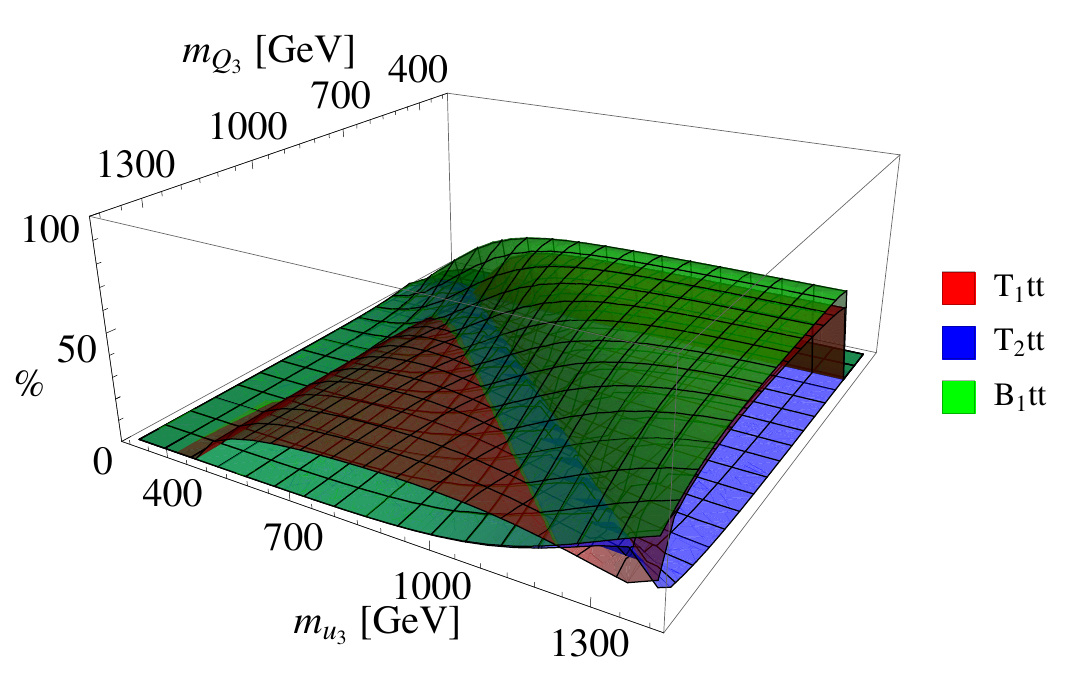}
  \includegraphics[width=0.44\textwidth]{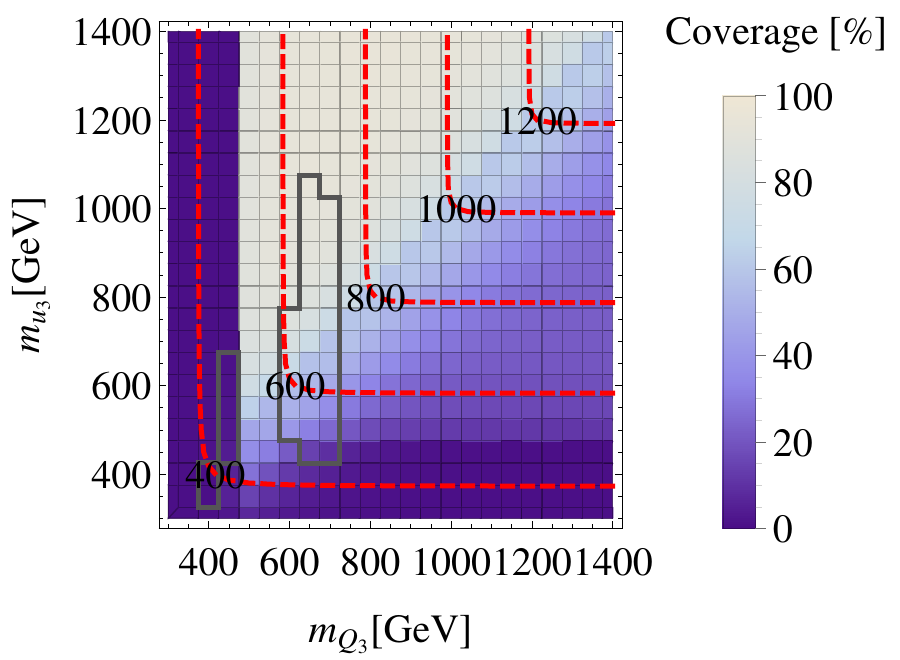}

\vspace{0mm}

  \includegraphics[width=0.44\textwidth]{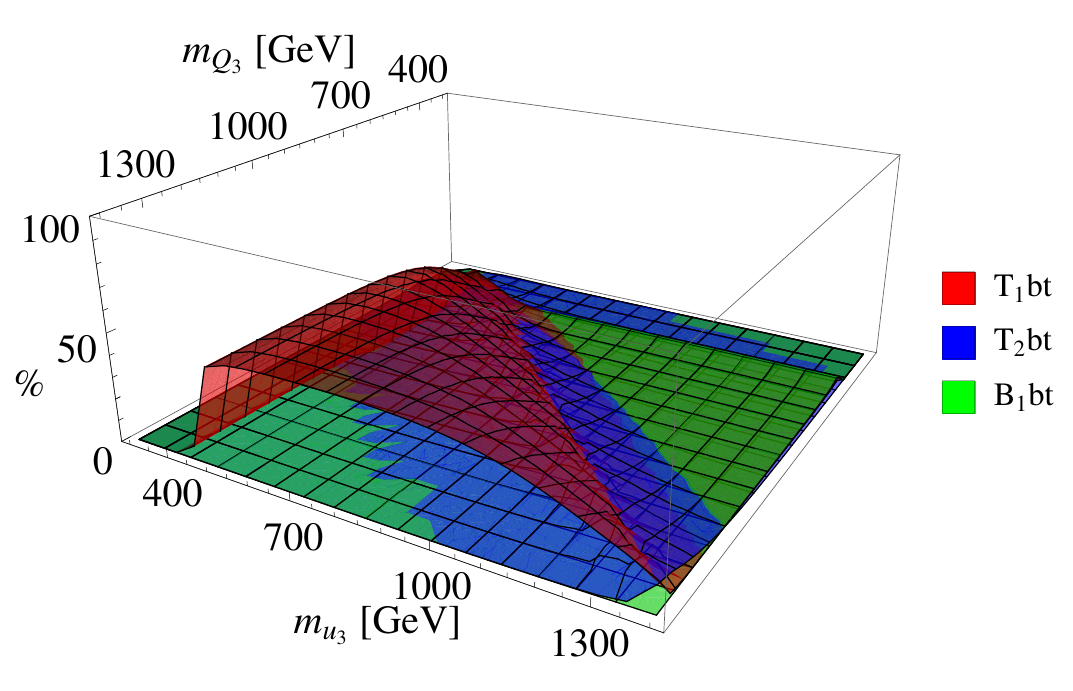}
  \includegraphics[width=0.44\textwidth]{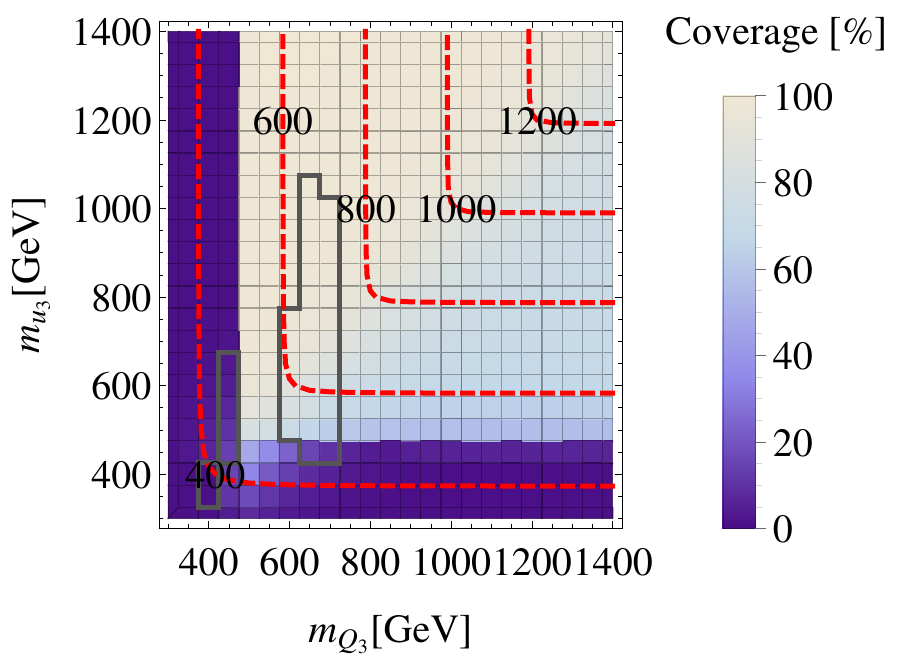}

\vspace{0mm}

  \includegraphics[width=0.44\textwidth]{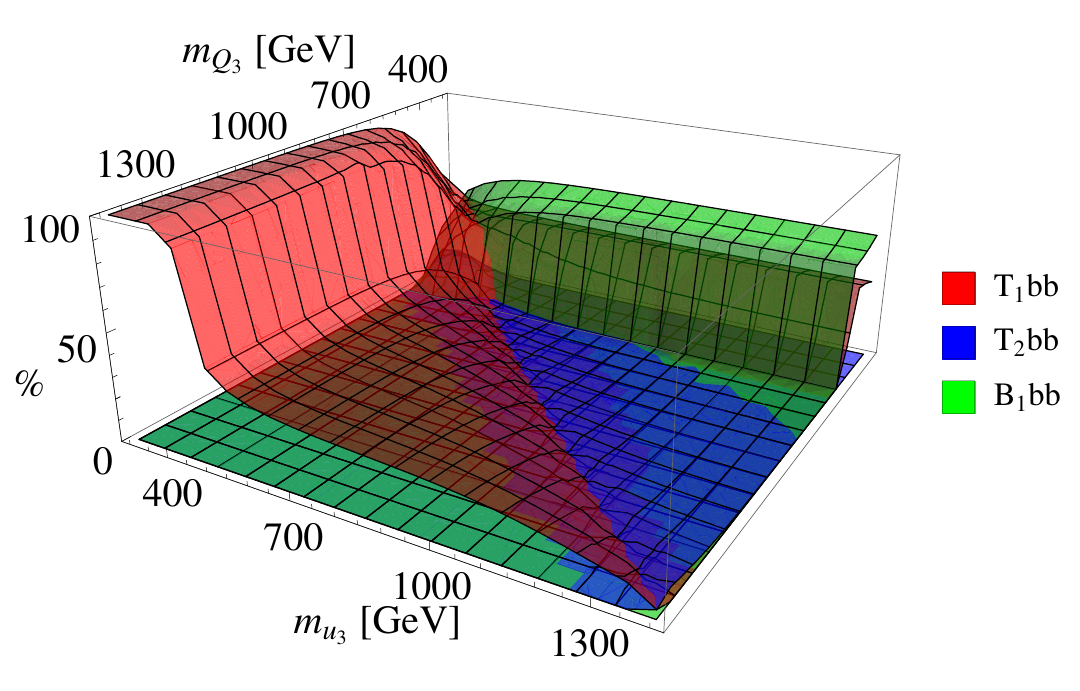}
  \includegraphics[width=0.44\textwidth]{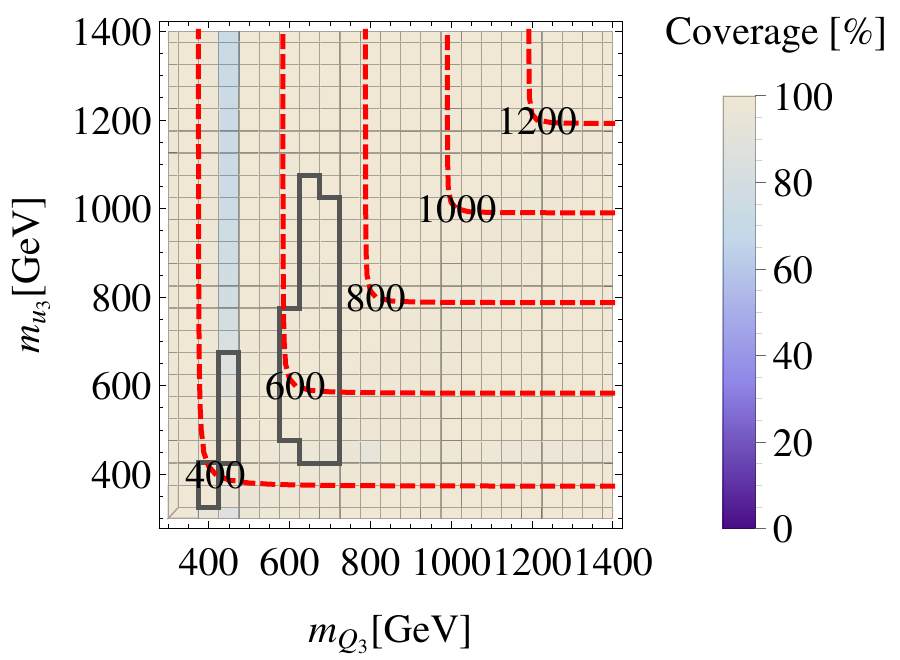}
  \caption{Left panels: The relative contribution of the different processes to the
    $\sigma_\tn{SUSY }$, i.e.~in red \mbox{$100\cdot\sigma_{T_1q_1q_2}/\sigma_{SUSY}$} as a function
    of $m_{Q_3}$ and $m_{u_3}$, where $T_1 q_1 q_2$ refers to $\tilde t_1$ pair production followed
    by a decay into a higgsino plus $q_1$ and $q_2$, respectively.  Right panels: Coverage, i.e.~sum
    of the relative contributions of the considered processes to $\sigma_\tn{SUSY }$. From top to
    bottom the considered channels are missing energy plus only $tt$ final states, $tt$ and $bt$
    final states, and $tt$, $bt$ and $bb$ final states. Note the different color scale of the lower
    right plot. The areas enclosed by the gray dashed lines show the points that are already
    excluded, determined by fastlim. Red dashed lines indicate the mass of the lightest stop in
    $\GeV$.}
  \label{fig:coverage_tt}
\end{figure}

Since the squark production cross section only depends on the squark mass and the known branching
ratios we can now determine which event topologies are the most dominant. In the left column of
Fig.~\ref{fig:coverage_tt} we show for $\mu=300\,\GeV$ the relative contribution to the total SUSY
cross section, defined as the sum of the squark pair production cross sections
$\sigma_\tn{SUSY}\equiv\sum_{S=\tilde t_1, \tilde t_2, \tilde b_1}\sigma_{S\bar S}$.  In the right
panels we show in color code the coverage defined as the sum of these relative contributions. The
larger the coverage the more of the signal cross section can be captured by looking into these
channels. Clearly considering only the decay of both squarks to top quarks and higgsinos is not
enough as the parameter space with $\mathcal{Q}<0$ is kinematically not accessible. Moreover in the
$m_{Q_3}>m_{u_3}$ half of the space this final state misses large parts of the signal since the
lighter stop decays dominantly into bottom quarks and charginos.\par
In the case where all three final states are taken into account, the coverage is nearly 100\%
throughout the parameter space, except along the line $m_{\tilde t_1}\approx m_t+m_{\tilde h}$ where
the top decay channel opens up. There it drops to 70--80\%, because in this narrow region also the
direct decay to a $W$ boson $\tilde t_1\to W+b+\widetilde \chi^0$ has a significant branching ratio.

For all parameter points we generate events for each of the up to nine signal processes using
\texttt{MadGraph5\_aMC@NLO}, version 2.1.1 \cite{Alwall:2014hca} at a center of mass energy
$\sqrt{s}=13\,\TeV$. No cuts are applied at the generator level. The matching up to two jets is done
with the MLM method in the shower-$k_T$ scheme \cite{Mangano:2001xp,Hoche:2006ph} with
\texttt{PYTHIA} version 6.426 \cite{Sjostrand:2006za}. We set the matching and the matrix element
cutoff scale to the same value of $m_{S}/6$ where $m_S$ is the mass of the produced squark.  We
checked and found that the differential jet distributions \cite{Lenzi:2009fi} are smooth with this
scale choice. The cross section for the signal processes is eventually rescaled by the NLO QCD and
NLL K-factors obtained from \texttt{NLL-fast}, version 3.0
\cite{Beenakker:1997ut,Beenakker:2010nq,Beenakker:2011fu}.

\subsection{Background sample}
\label{sec:background-sample}

In our analysis we use top tagging methods based on jet substructure techniques. We therefore focus
on the decay of the squarks into a neutralino and a bottom quark or a hadronically decaying top
quark. The latter will generate between one to three distinct jets and the former will generate
missing energy. Our final state therefore consists of missing energy and up to six jets. As
background we thus consider the following four processes, all generated with
\texttt{MadGraph5\_aMC@NLO} version 2.1.1 \cite{Alwall:2014hca} and showered with \texttt{PYTHIA}
version 6.426 \cite{Sjostrand:2006za}.
\begin{itemize}
\item $\boldsymbol{Wj}${\bf :} $p p \to W_\ell + (2+X)\tn{jets}$, where we merge up to 4 jets in the
  five flavor scheme and demand that the $W$ decays into leptons (including taus), such that the
  neutrino accounts for the missing energy.
\item $\boldsymbol{Zj}${\bf :} $p p \to Z_\nu + (2+X)\tn{jets}$, where we merge up to 4 jets in the
  five flavor scheme and the $Z$ decays into two neutrinos and hence generates missing energy. In
  both channels $Wj$ and $Zj$ we demand missing transverse energy of at least $70\,\GeV$ at the
  generator level.
\item $\boldsymbol{Zt\bar t}${\bf :} $p p \to Z_\nu + t + \bar t$, where both top quarks decay
  hadronically, faking the top quarks from the squark decay and the $Z$ decays again into two
  neutrinos to generate missing energy. This cross section is known at NLO QCD \cite{Kardos:2011na}
  and rescaled by the corresponding K-factor.
\item $\boldsymbol{t\bar t}${\bf :} $p p \to t\bar t + \tn{jets}$, where one top decays hadronically
  and the other one leptonically to emit a neutrino, which accounts for missing energy. The
  NNLO+NNLL QCD K-factor is obtained from \texttt{Top++} version 2.0 \cite{Czakon:2013goa} and
  multiplied with the cross section.
\end{itemize}

\section{Analysis}
\label{sec:analysis-1}

\subsection{Reconstruction}
\label{sec:reconstruction}

For the reconstruction of the events we use \texttt{ATOM} \cite{ATOM:20xxaaa}, based on
\texttt{Rivet} \cite{Buckley:2010ar}. Electrons and muons are reconstructed if their transverse
momentum is greater than $10\,\GeV$ and their pseudo-rapidity is within $\abs{\eta}<2.47$ for
electrons and $\abs{\eta}<2.4$ for muons. Jets for the basic reconstruction are clustered with
\texttt{FastJet} version 3.1.0 \cite{Cacciari:2011ma} with the anti-$k_t$ algorithm
\cite{Cacciari:2008gp} and a jet radius of 0.4\,. Only jets with $p_T>20\,\GeV$ and with
$\abs{\eta}<2.5$ are kept. For the overlap removal we first reject jets that are within
$\Delta R=0.2$ of a reconstructed electron and then all leptons that are within $\Delta R=0.4$ of
one of the remaining jets. All constituents of the clustered jets are used as input for the
following re-clustering as described below.\par
The underlying idea behind the reconstruction described in the following is to cover a large range
of possible boosts of the top quark. We therefore gradually increase the cluster radius and employ
successively both the HEPTop and the BDRS tagger. This allows us to reduce background significantly
while maintaining a high signal efficiency.\par
\begin{figure}[tb]
  \centering
  \includegraphics[width=0.85\textwidth]{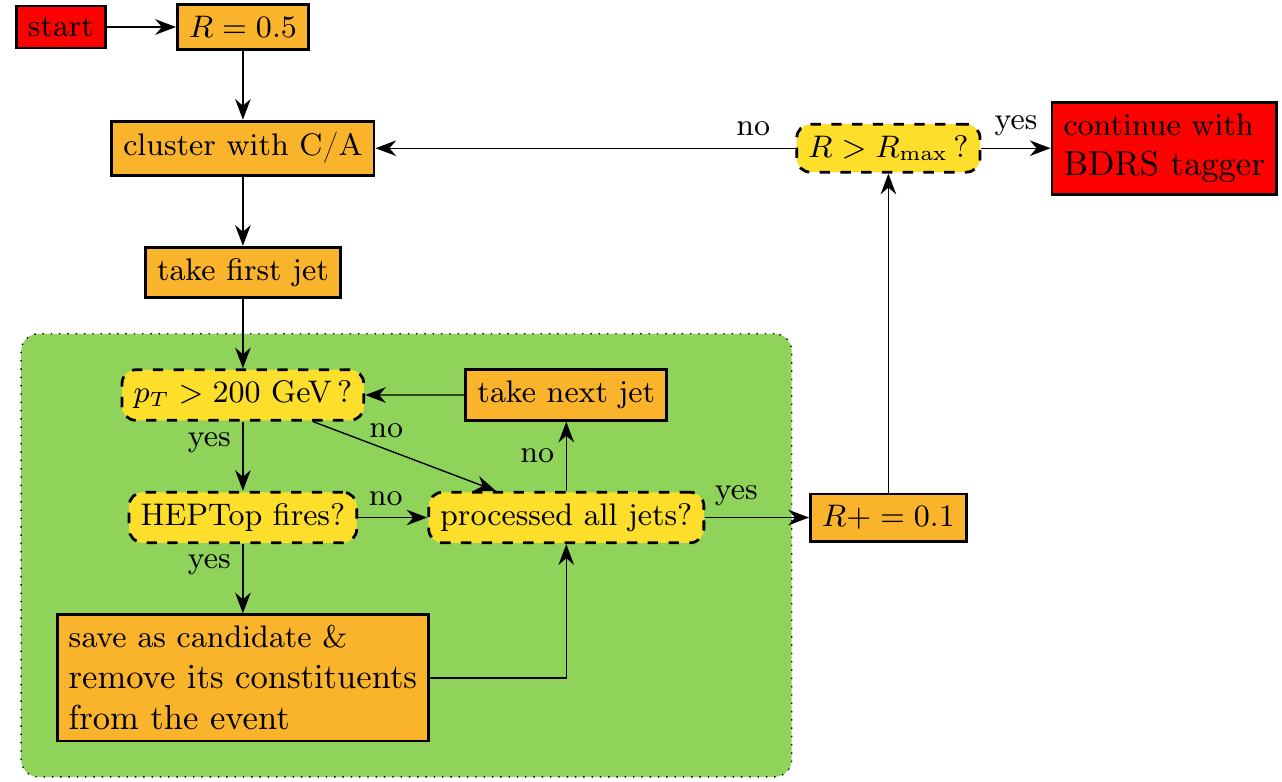}
  \caption{Flowchart of the top reconstruction with the HEPTopTagger.}
  \label{fig:HEPTop}
\end{figure}

A flowchart for the reconstruction of the top candidates with the HEPTopTagger is shown in
Fig.~\ref{fig:HEPTop}. First the cluster radius is set to $R=0.5$ and the constituents of the
initial anti-$k_t$ jets are re-clustered with the Cambridge-Aachen (C/A) algorithm
\cite{Dokshitzer:1997in,Wobisch:1998wt}. Then for each of the obtained jets we check if its
transverse momentum is greater than $200\,\GeV$ and if the HEPTopTagger tags it as a top. In this
case we save it as candidate for a signal final state and remove its constituents from the event
before moving on to the next jet. Once all jets are analyzed as described above we increase the
cluster radius by 0.1 and start over again with re-clustering the remaining constituents of the
event. This loop continues until we exceed the maximal clustering radius of $R_\tn{max}=1.5$\,.\par
\begin{figure}[tb!]
  \centering
  \includegraphics[width=0.95\textwidth]{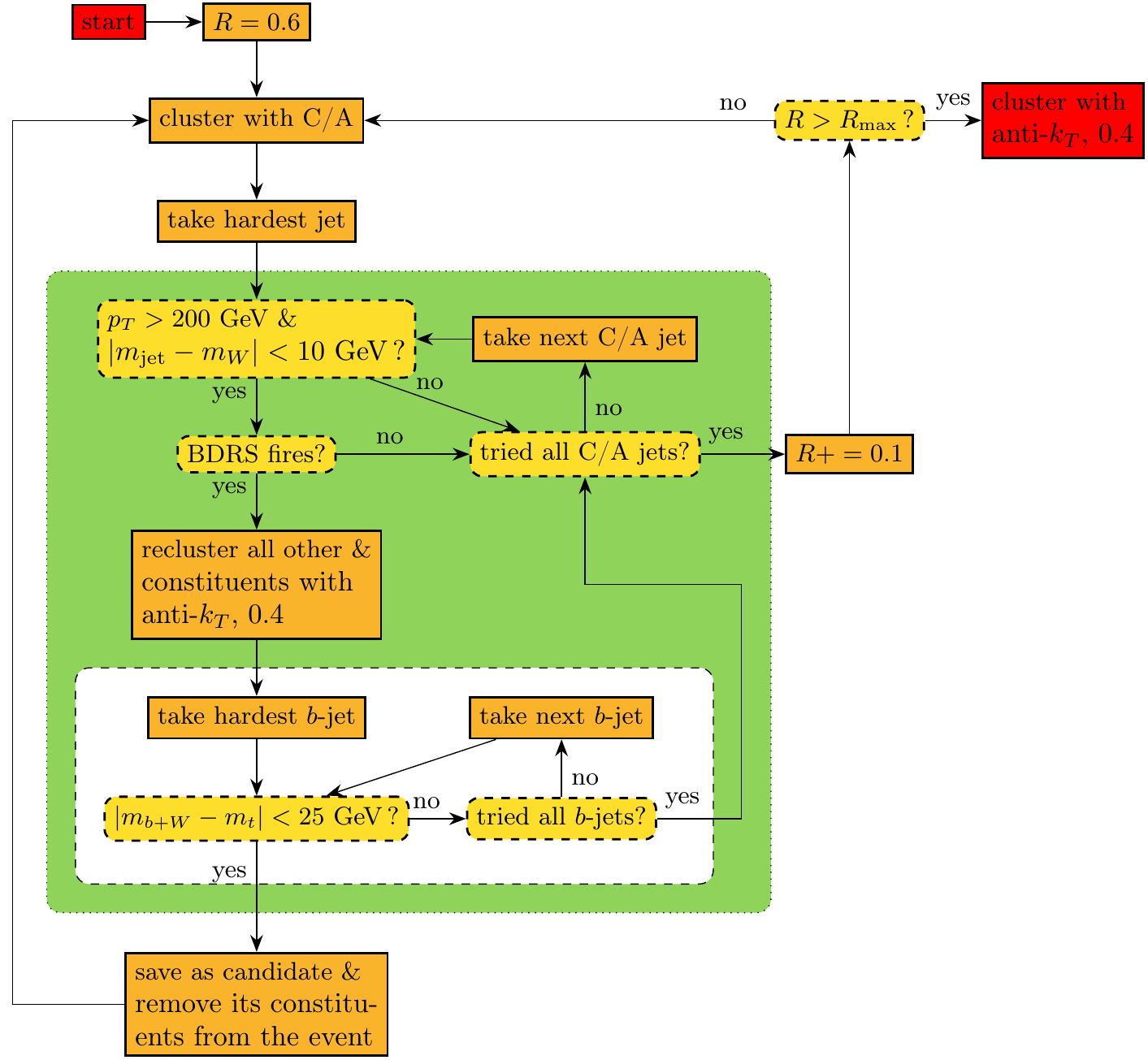}
  \caption{Flowchart of the top reconstruction with the BDRS Tagger.}
  \label{fig:BDRS}
\end{figure}
After the reconstruction with the HEPTopTagger is finished we continue the reconstruction of the top
candidates with the BDRS tagger as sketched in Fig.~\ref{fig:BDRS}. We choose our initial cluster
radius $R=0.6$ and cluster the remaining constituents of the event with the Cambridge-Aachen
algorithm. Since we now only expect to find $W$ candidates with the BDRS Tagger and need to combine
them with a $b$-jet to form top candidates the order in which we analyze the jets is no longer
arbitrary. Starting with the hardest C/A jet we check if its transverse momentum exceeds
$200\,\GeV$, its invariant mass is within $10\,\GeV$ of the $W$ mass, and the BDRS Tagger recognizes
a mass drop. In the case that one of the above requirements fails we proceed with the next hardest
C/A jet until either one jet fulfills them or we find all jets to fail them. In the latter case we
increase the cluster radius by 0.1 and repeat the C/A clustering and analyzing of jets until the
radius gets greater than $R_\tn{max}=1.5$. Once a jet fulfills all the previous criteria we need to
find a $b$-jet to create a top candidate. To do this we recluster the constituents of the event that
are not part of the given jet with the anti-$k_T$ algorithm and a cone radius of 0.4 and pass them
on to the $b$-tagger\footnote{For the $b$-tagger we mimic a tagger with efficiency 0.7 and rejection
  50. We check if a given jet contains a bottom quark in its history and tag it as $b$-jet with a
  probability of 70\,\% if this is the case and with a probability of 2\,\% otherwise. Since the
  same jet may be sent to the $b$-tagger at different stages of the reconstruction process we keep
  the results of the $b$-tagger in the memory and reuse them each time it gets a previously analyzed
  jet. This way we avoid assigning different tagging results to the same jet.}. Starting with the
hardest $b$-jet we check if the combined invariant mass of the $W$ candidate and the $b$-jet is
within $25\,\GeV$ of the top quark mass. If such a combination is found it is saved as a candidate
and its constituents are removed from the event. The remaining constituents of the event are
reclustered with the C/A algorithm and the procedure repeats. Alternatively, if all $b$-jets fail to
produce a suitable top candidate the next C/A jet is analyzed. Once the C/A cluster radius exceeds
$R_\tn{max}$ the remaining constituents of the event are clustered with the anti-$k_T$ algorithm
with radius 0.4 and passed on to the $b$-tagger. Those that get $b$-tagged are saved as candidates
of the signal final state as well.

\subsection{Analysis cuts}
\label{sec:analysis-cuts}

\begin{figure}[tb]
  \centering
    \subfloat[Before cut \#1]{\includegraphics[width=0.32\textwidth]{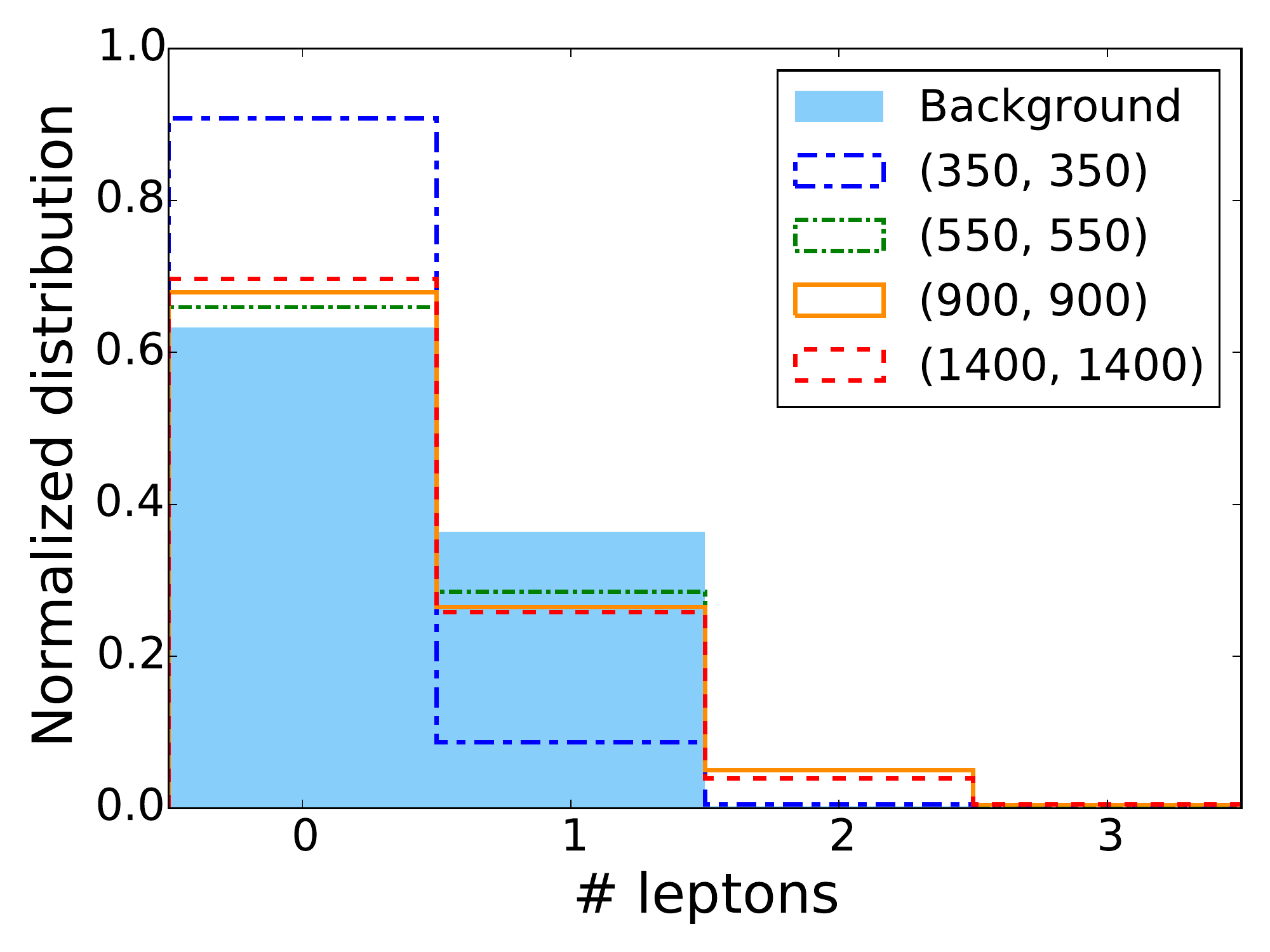}\label{fig:NLeptons}}
    \subfloat[Before cut \#2]{\includegraphics[width=0.32\textwidth]{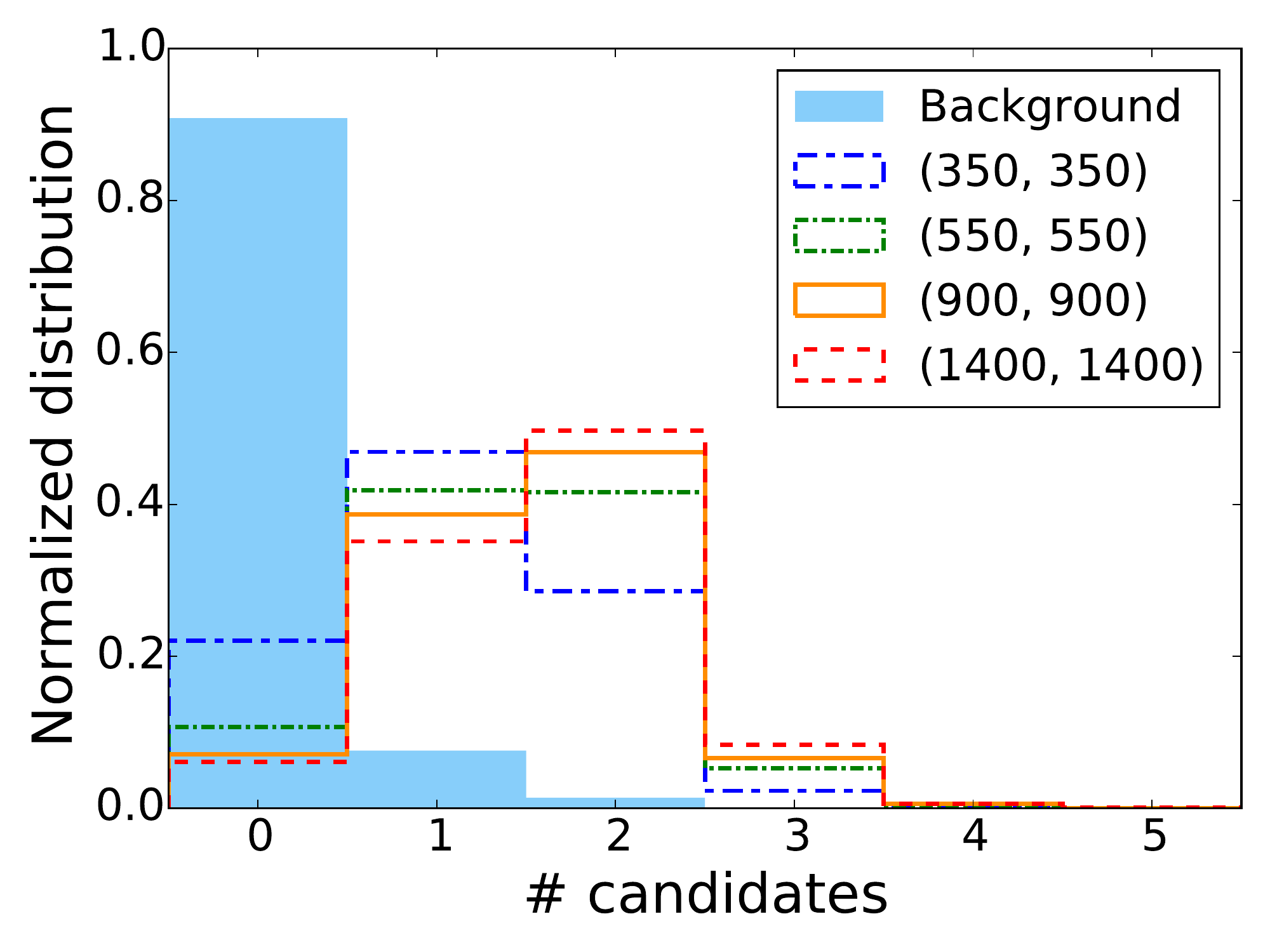}\label{fig:NCand}}
    \subfloat[Before cut \#3]{\includegraphics[width=0.32\textwidth]{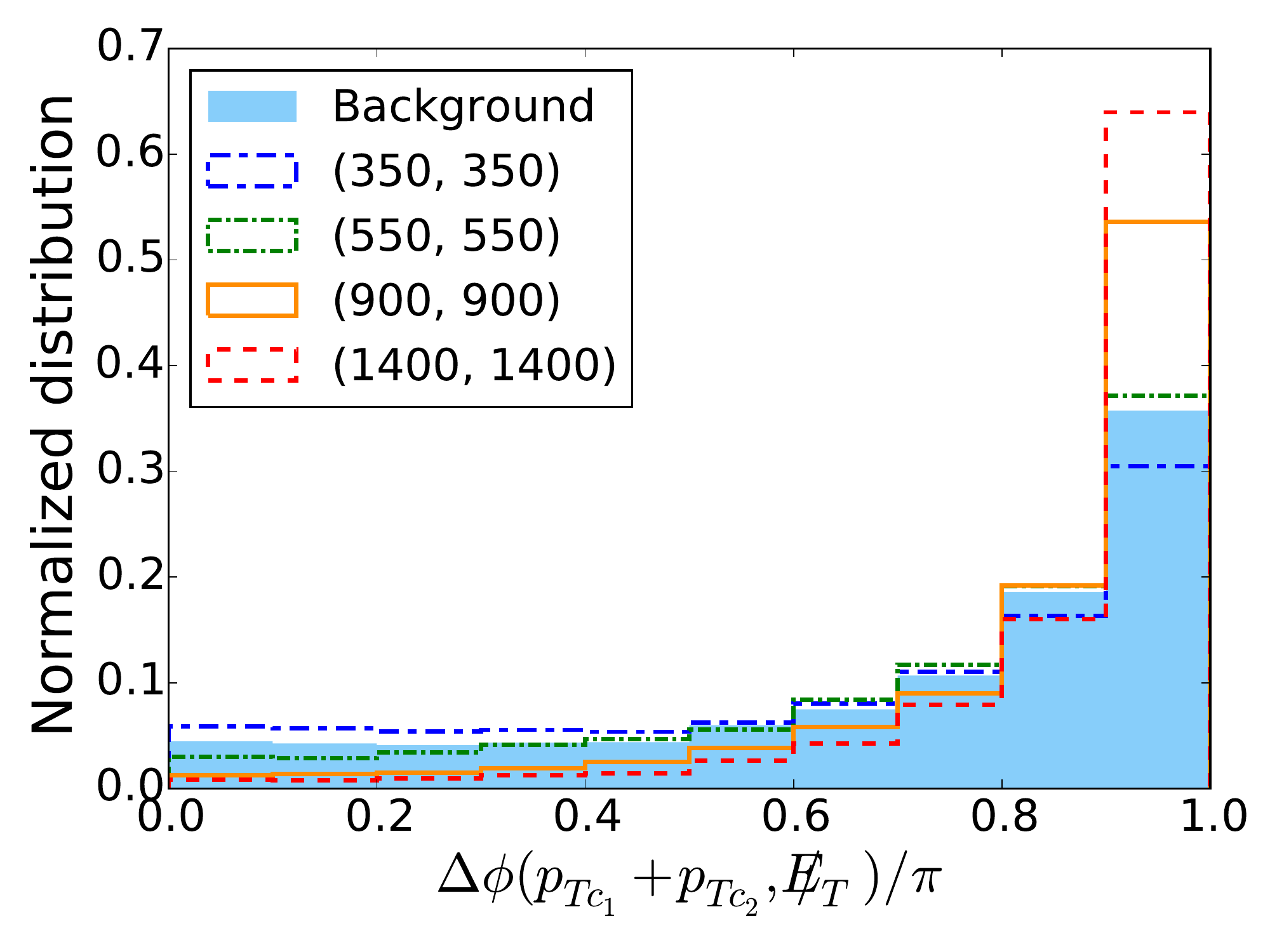}\label{fig:dphi12E}}\\
    \subfloat[Before cut \#4]{\includegraphics[width=0.32\textwidth]{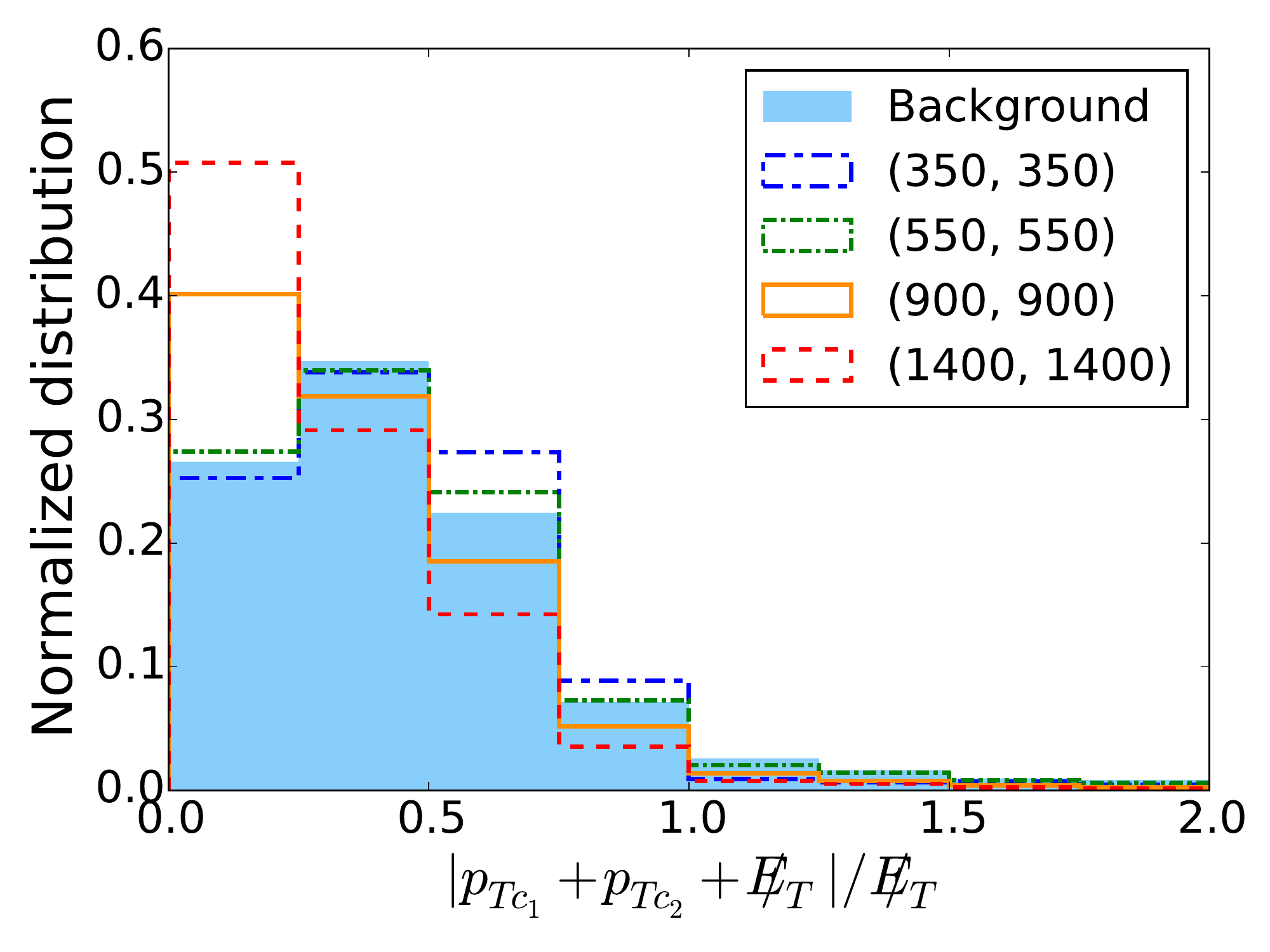}\label{fig:pTsum}}
    \subfloat[Before cut \#5]{\includegraphics[width=0.32\textwidth]{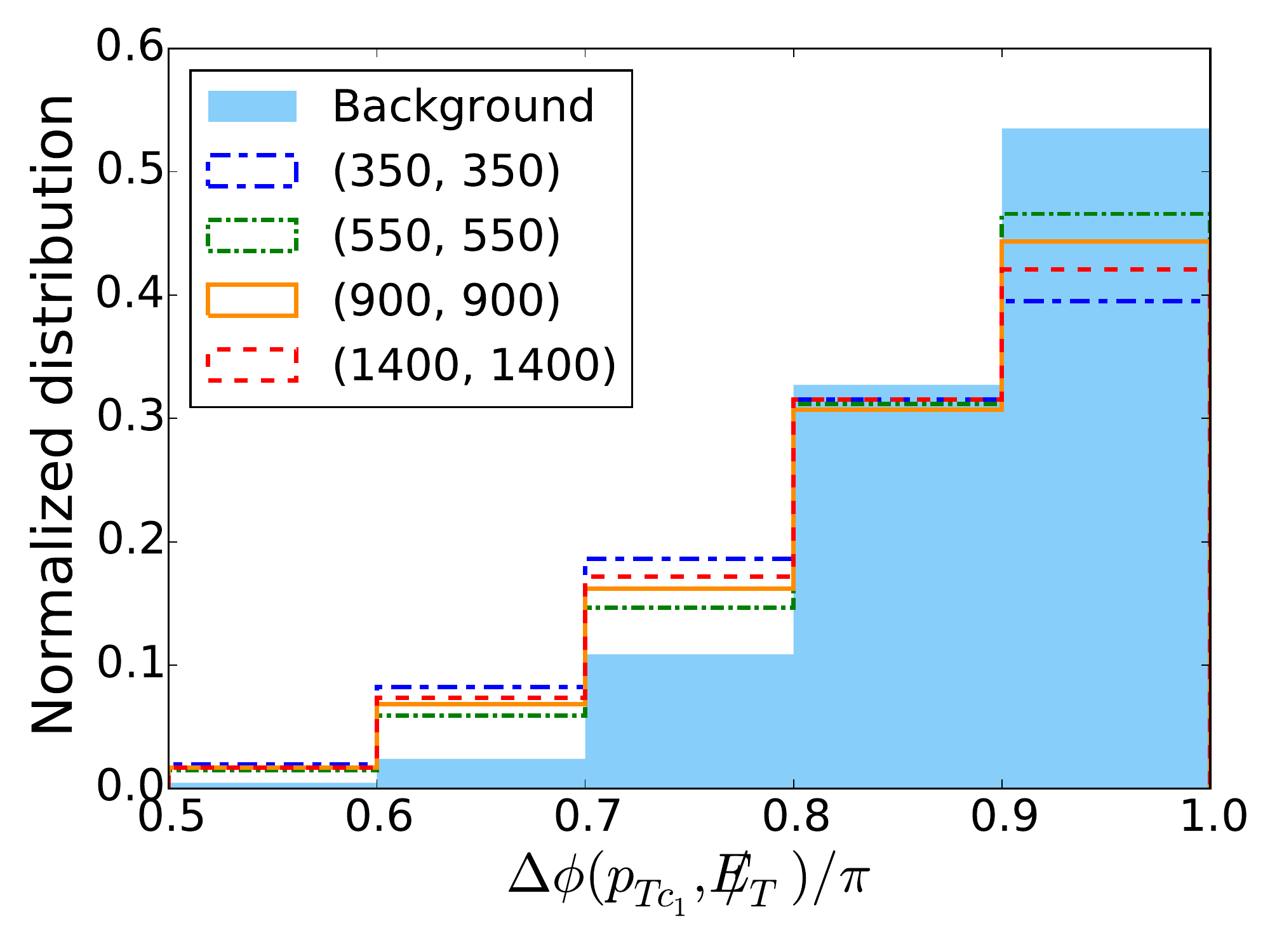}\label{fig:c1MET}}
    \subfloat[Before cut \#6]{\includegraphics[width=0.32\textwidth]{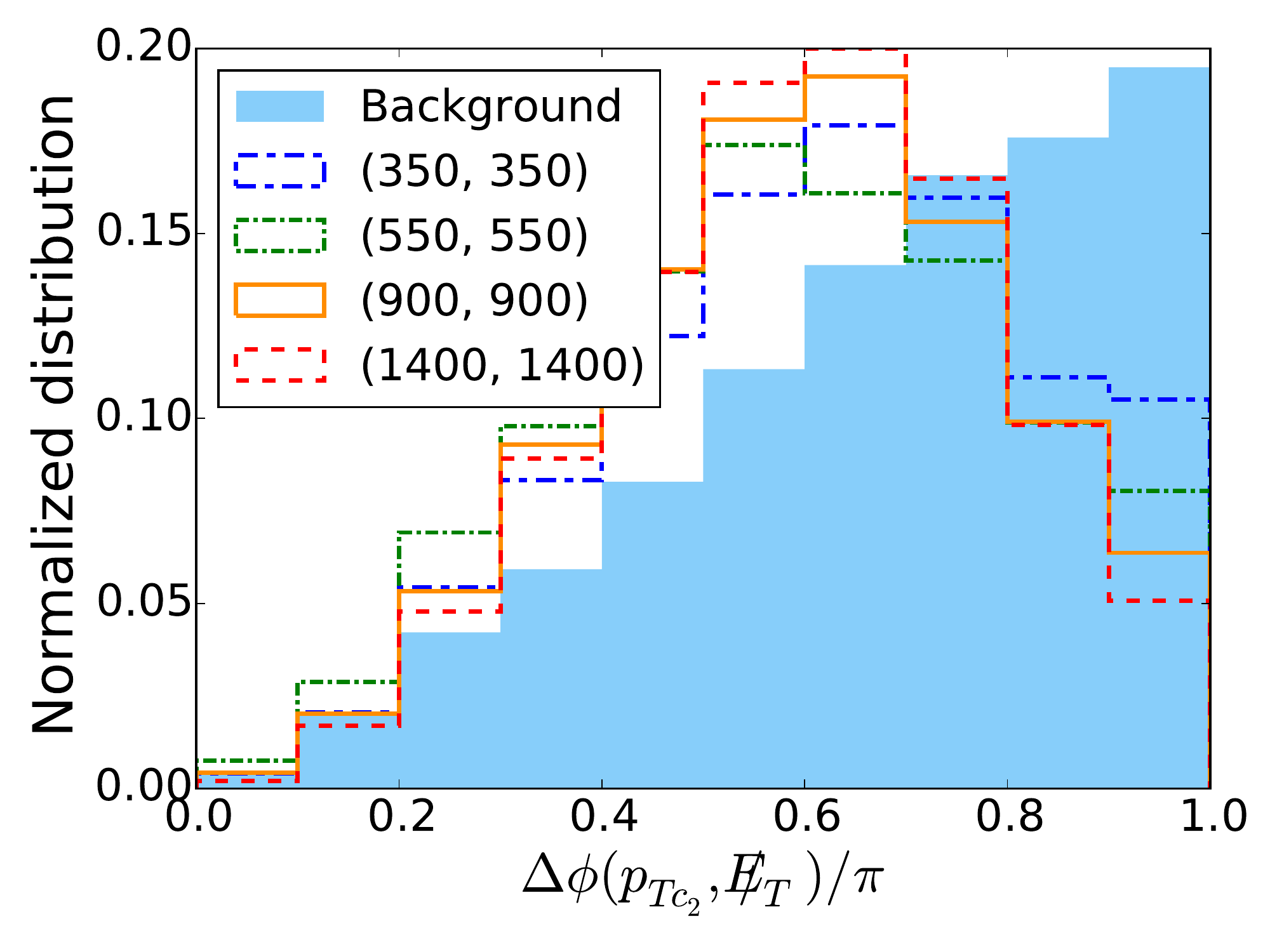}\label{fig:c2MET}}
    \caption{Normalized distributions of the background and signal processes before the cut on the
      respective observable. The numbers in the brackets stand for the soft mass parameters
      $m_{Q_3}$ and $m_{u_3}$ in GeV and the sums in Subfigs.~\protect\subref{fig:dphi12E} and
      \protect\subref{fig:pTsum} are understood to be vectorial. In all plotted samples
      $\mu=300\,\GeV$.}
  \label{fig:NormedDistributions}
\end{figure}

After having reconstructed the candidates for the hadronic final states of the signal --- top
candidates and $b$-tagged anti-$k_T$ jets --- we proceed with the analysis cuts. As our premise is to
make a scale invariant analysis, we must avoid to introduce scales through the cuts. We propose the
following ones and show the respective distribution before each cut in Fig.~\ref{fig:NormedDistributions}.
\begin{enumerate}
\item {\bf Zero leptons:} The leptons or other particles that are emitted by the decaying chargino
  or second lightest neutralino are too soft to be seen by the detector. Moreover, since we focus on
  the hadronic decay modes, no leptons should be present in the signal events. In the Background
  however, they are produced in the leptonic decays which are necessary to generate missing
  energy. We therefore demand zero reconstructed electrons or muons.
\item {\bf Exactly two candidates:} The visible part of the signal process consists of two hadronic
  final states as defined above. In the rare case that an event contains more but in particular in
  the cases where an event contains less than these two candidates it is rejected. This means that
  no $b$-jets beyond possible $b$ candidates are allowed.
\item {\bf $\boldsymbol{\Delta\phi(\vec{p}_{T,c_1}+\vec{p}_{T,c_2},\slashed{\vec{E}}_T)>0.8\pi}$:}
  Since we cannot determine the two neutralino momenta individually, it is impossible to reconstruct
  the momenta of the initial squarks. Yet, we can make use of the total event shape. In the signal,
  the transverse missing energy is the combination of the two neutrino momenta and therefore
  balances the transverse momenta of the two candidates. Consequently the vectorial sum of the
  candidate's transverse momenta $\vec{p}_{T,c_i}$ has to point in the opposite direction of the missing energy.
\item {\bf
    $\boldsymbol{\abs{\vec{p}_{T,c_1}+\vec{p}_{T,c_2}+\vec{\slashed{E}}_T}/\slashed{E}_T<0.5}$: }
  This cut is based on the same reasoning as the previous one. The absolute value of the summed
  candidate's transverse momenta and the missing transverse energy needs to be small. In order to
  maintain scale invariance we normalize the result by $\slashed{E}_T$.
\item {\bf $\boldsymbol{\Delta\phi(\vec{p}_{T,c_1}, \slashed{\vec{E}}_T)<0.9\pi}$:} By this cut we
  demand that the missing transverse energy and the transverse momentum of the harder of the two
  candidates are not back-to-back. Since the two produced squarks are of the same type and the
  higgsinos are mass degenerate, the recoil of the top or bottom quarks against the respective
  higgsino will be the same in the squark rest-frame. Therefore, the two neutralinos should
  contribute about equally to the missing energy and spoil the back-to-back orientation that is
  present for each top neutralino pair individually. We therefore reject events where one top
  candidate recoils against an invisible particle and the second candidate does not. Moreover we can
  thus reject events where the missing energy comes from a mismeasurement of the jet momentum.
\item {\bf $\boldsymbol{\Delta\phi(\vec{p}_{T,c_2}, \slashed{\vec{E}}_T)<0.8\pi}$:} This cut
  exploits the same reasoning as the previous one.
\end{enumerate}

\begin{figure}[tb]
  \centering
  \includegraphics[width=\textwidth]{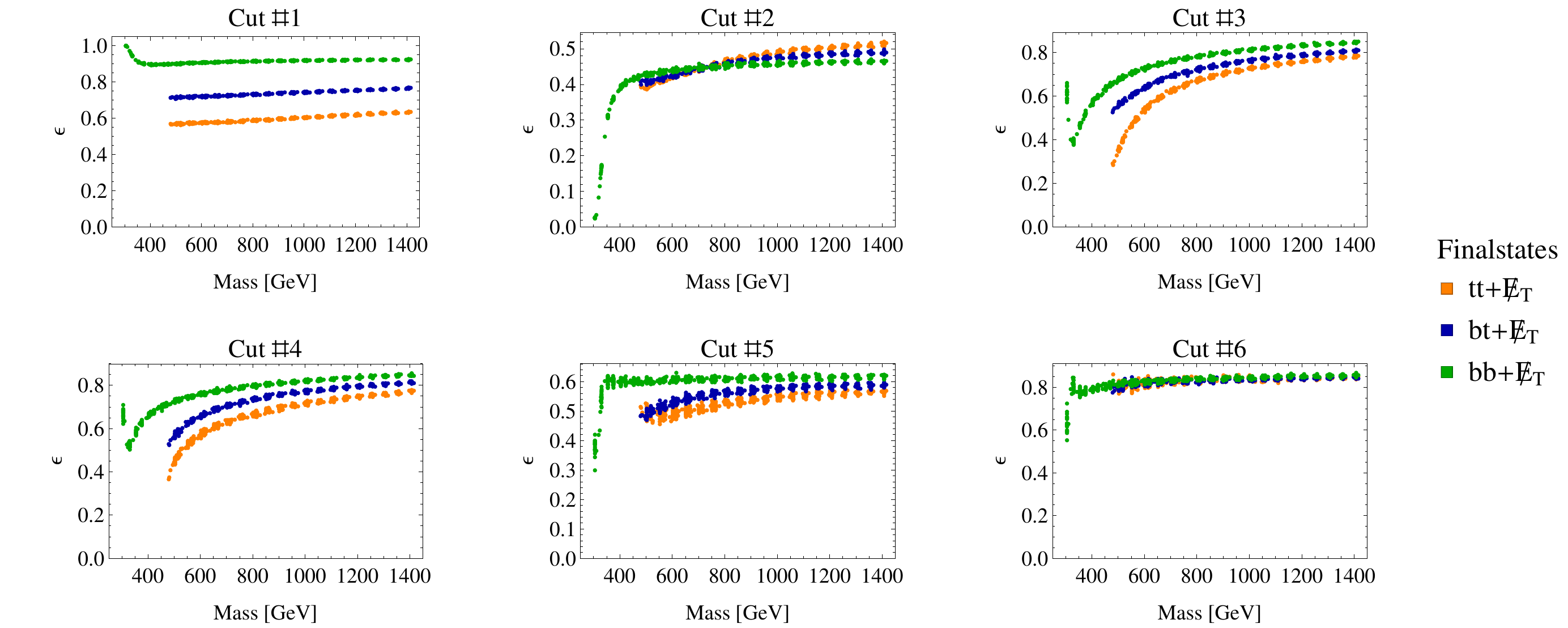}
  \caption{Efficiency of the cuts as a function of the squark mass for all samples with
    $\mu=300\,\GeV$.}
  \label{fig:Efficiency}
\end{figure}

\begin{figure}[!htbp]
  \subfloat
  {
    \begin{minipage}[t!]{0.46\linewidth}
      \centering
      \includegraphics[width=1.\textwidth]{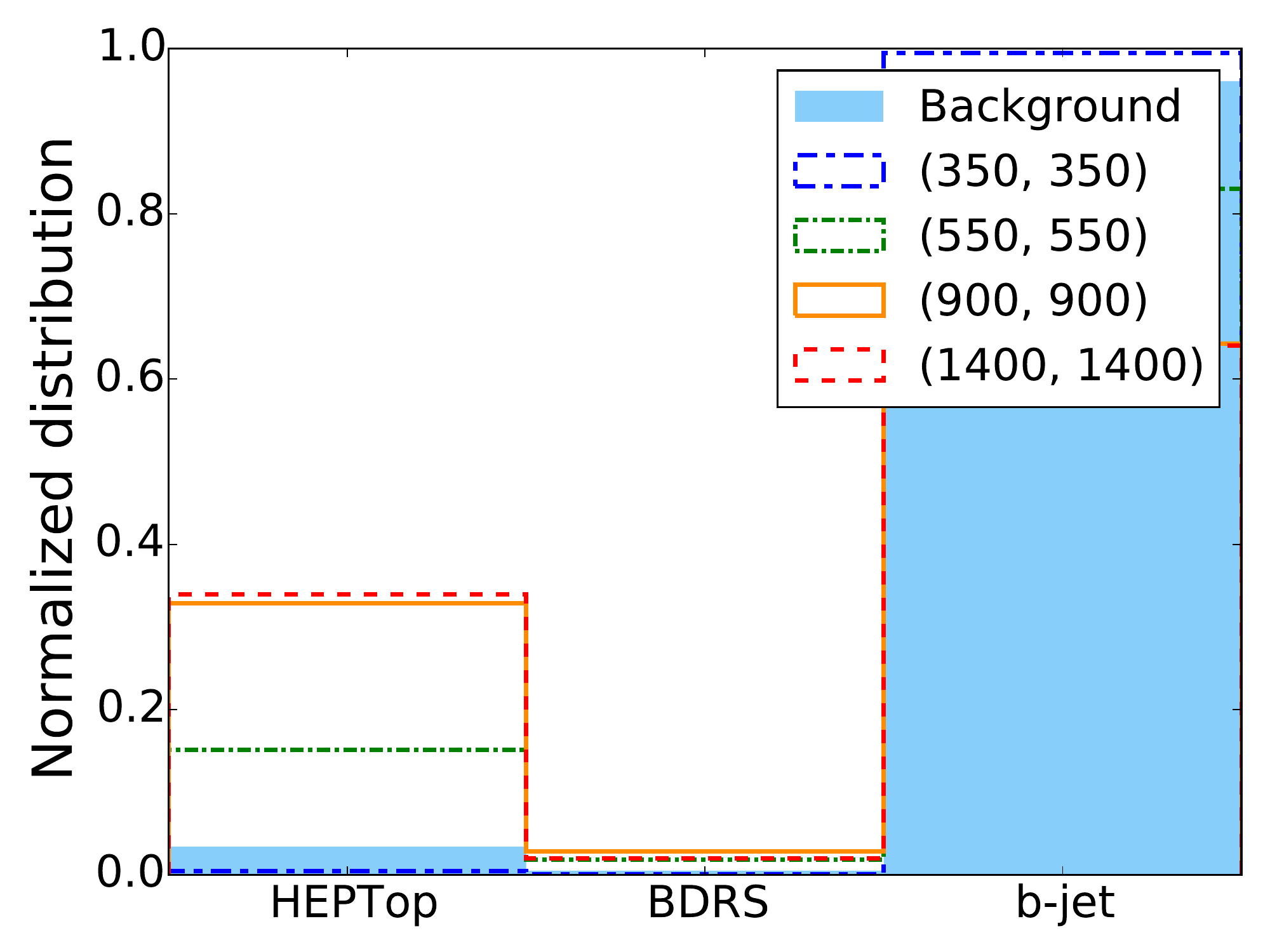}\label{fig:kind}
    \end{minipage}}
  \subfloat
  {
    \begin{minipage}[t!]{0.5\linewidth}
      \centering
      \rule[7pt]{0pt}{2pt}
      \includegraphics[width=\textwidth]{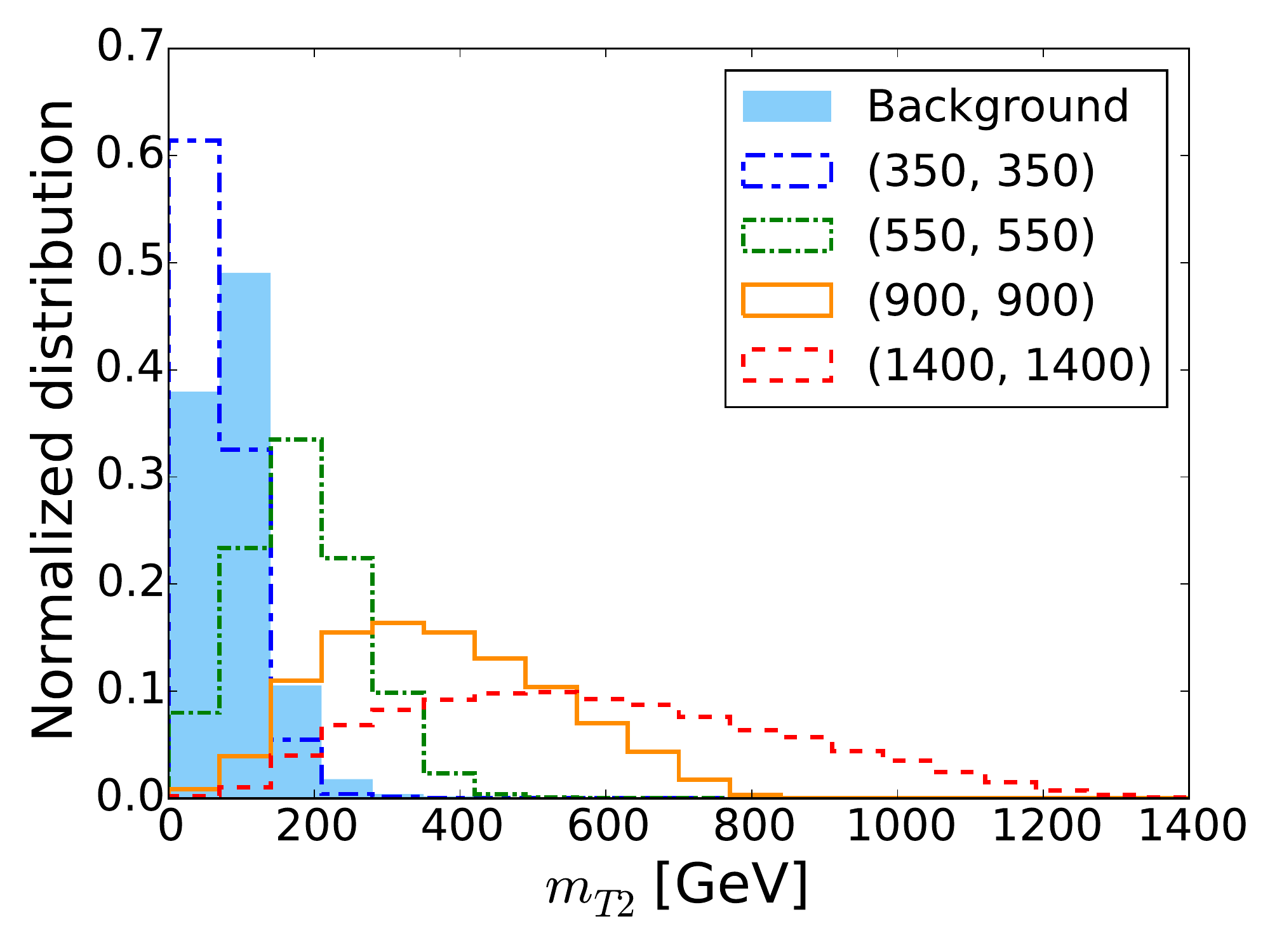}\label{fig:mt2}
    \end{minipage}}
  \caption{Normalized distribution of the type of tags of the candidates and of $m_{T2}$ after the
    last cut.}
  \label{fig:NormedDistributions2}
\end{figure}

\begin{figure}[tb]
  \centering
  \includegraphics[width=\textwidth, trim=0cm 0cm 0cm 5.5cm, clip=true]{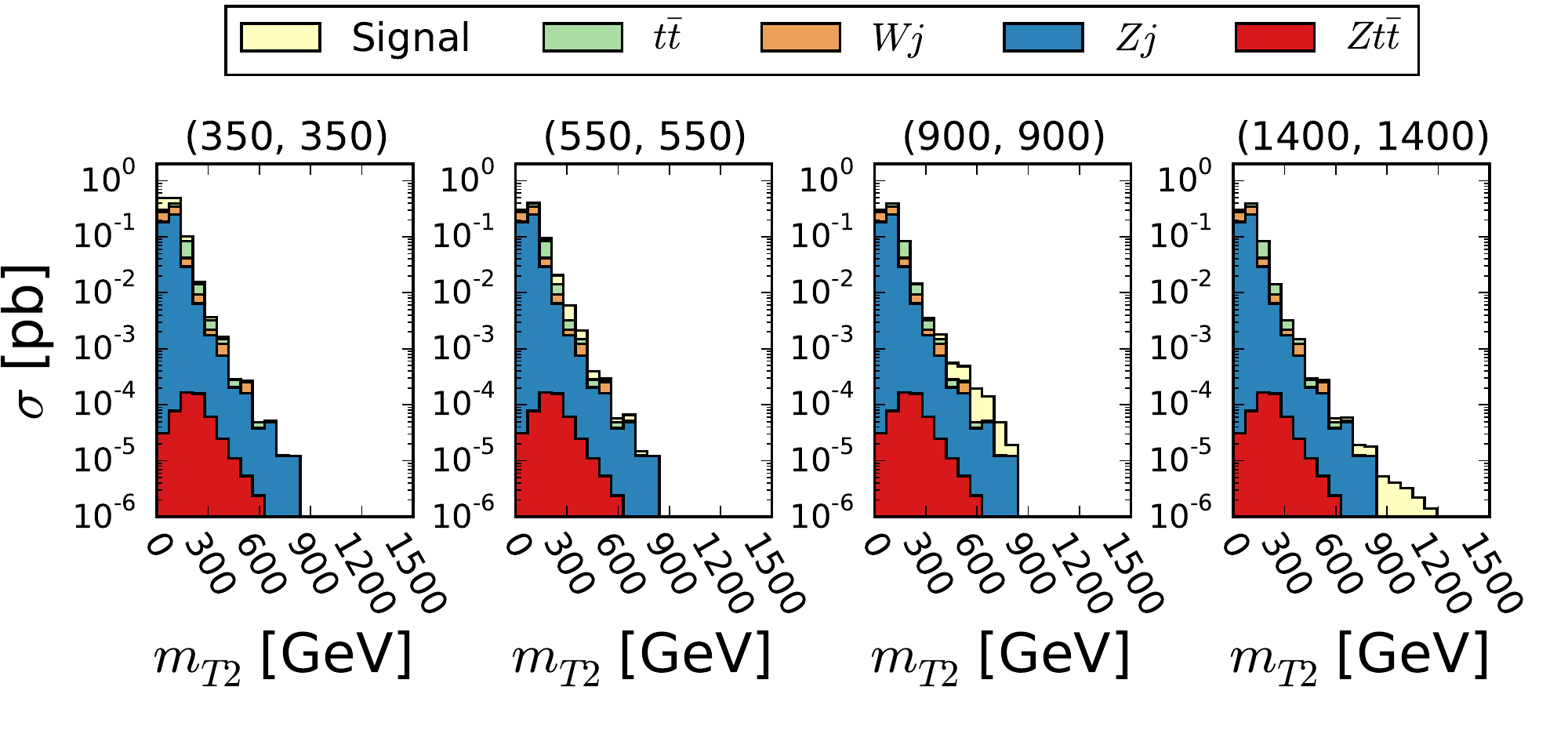}
  \caption{Stacked distribution of $m_{T2}$ after all cuts. The plots are based on the samples with
    $\mu=300\,\GeV$ and the numbers in the brackets refer to the soft mass parameters $m_{Q_3}$ and
    $m_{u_3}$ in GeV.}
  \label{fig:StackedMT2}
\end{figure}

\begin{table}[tb]
  \footnotesize
  \centering
  \begin{tabular}[]{|l|r|r|r|r|}\hline
    cut  & $Zt\bar t$ & $Zj$ & $Wj$  & $t\bar t$\\\hline
    0. no cut        &$1.13\cdot10^{-2}$ &$2.11\cdot10^{2} $ &$5.38\cdot10^{2} $ &$1.94\cdot10^{1}$ \\\hline
    1. 0 leptons     &$1.01\cdot10^{-2}$ &$2.09\cdot10^{2} $ &$2.71\cdot10^{2} $ &$6.32\cdot10^{0}$ \\\hline
    2. 2 candidates  &$4.04\cdot10^{-3}$ &$3.27\cdot10^{0}$ &$2.62\cdot10^{0}$ &$2.34\cdot10^{0}$ \\\hline
    3. $\Delta\phi(\boldsymbol{p_{T,c_1}}+\boldsymbol{p_{T,c_2}},
    \boldsymbol{\slashed{E}_T})>0.8\pi$
                     &$2.29\cdot10^{-3}$ &$2.31\cdot10^{0}$ &$1.48\cdot10^{0}$ &$8.19\cdot10^{-1}$ \\\hline
    4. $\abs{\boldsymbol{p_{T,c_1}}+\boldsymbol{p_{T,c_2}}+\boldsymbol{\slashed{E}_T}}/\slashed{E}_T<0.5$
                     &$1.18\cdot10^{-3}$ &$1.62\cdot10^{0}$ &$8.18\cdot10^{-1}$ &$3.22\cdot10^{-1}$ \\\hline
    5. $\Delta\phi(\boldsymbol{p_{T,c_1}},\boldsymbol{\slashed{E}_T})<0.9\pi$
                     &$6.12\cdot10^{-4}$ &$7.46\cdot10^{-1}$ &$3.74\cdot10^{-1}$ &$1.62\cdot10^{-1}$ \\\hline
    6. $\Delta\phi(\boldsymbol{p_{T,c_2}},\boldsymbol{\slashed{E}_T})<0.8\pi$
                     &$5.37\cdot10^{-4}$ &$4.70\cdot10^{-1}$ &$2.07\cdot10^{-1}$ &$1.18\cdot10^{-1}$ \\\hline
  \end{tabular}
  \caption{Cutflow for the background processes. The numbers give the cross section in picobarns after the
    respective cut.}
  \label{tab:cutflowBKG}
\end{table}

\begin{table}[tb]
  \scriptsize
  \centering
  \begin{tabular}[]{|l|r|r|r|r|r|r|r|r|}\hline
    & \multicolumn{2}{c|}{(350,350)} & \multicolumn{2}{c|}{(550,550)} & \multicolumn{2}{c|}{(900,900)} & \multicolumn{2}{c|}{(1\,400,1\,400)}\\\hhline{~--------}
    cut No.& $\sigma$ & $S/B$ & $\sigma$ & $S/B$ & $\sigma$ & $S/B$ & $\sigma$ & $S/B$\\\hhline{|=|=|=|=|=|=|=|=|=|}
    0 \rule[-1pt]{0pt}{8pt}&$8.67\cdot10^{0}$ &$1.13\cdot10^{-2}$ &$6.57\cdot10^{-1}$ &$8.56\cdot10^{-4}$ &$2.69\cdot10^{-2}$ &$3.50\cdot10^{-5}$ &$8.43\cdot10^{-4}$ &$1.10\cdot10^{-6}$  \\\hline
    1 \rule[-1pt]{0pt}{8pt}&$7.88\cdot10^{0}$ &$1.62\cdot10^{-2}$ &$4.33\cdot10^{-1}$ &$8.91\cdot10^{-4}$ &$1.82\cdot10^{-2}$ &$3.73\cdot10^{-5}$ &$5.93\cdot10^{-4}$ &$1.22\cdot10^{-6}$  \\\hline
    2 \rule[-1pt]{0pt}{8pt}&$2.30\cdot10^{0}$ &$2.80\cdot10^{-1}$ &$1.82\cdot10^{-1}$ &$2.22\cdot10^{-2}$ &$8.53\cdot10^{-3}$ &$1.04\cdot10^{-3}$ &$2.93\cdot10^{-4}$ &$3.56\cdot10^{-5}$  \\\hline
    3 \rule[-1pt]{0pt}{8pt}&$1.10\cdot10^{0}$ &$2.39\cdot10^{-1}$ &$1.01\cdot10^{-1}$ &$2.20\cdot10^{-2}$ &$6.13\cdot10^{-3}$ &$1.33\cdot10^{-3}$ &$2.31\cdot10^{-4}$ &$5.02\cdot10^{-5}$  \\\hline
    4 \rule[-1pt]{0pt}{8pt}&$6.50\cdot10^{-1}$ &$2.36\cdot10^{-1}$ &$6.32\cdot10^{-2}$ &$2.29\cdot10^{-2}$ &$4.44\cdot10^{-3}$ &$1.61\cdot10^{-3}$ &$1.84\cdot10^{-4}$ &$6.69\cdot10^{-5}$  \\\hline
    5 \rule[-1pt]{0pt}{8pt}&$3.96\cdot10^{-1}$ &$3.09\cdot10^{-1}$ &$3.40\cdot10^{-2}$ &$2.65\cdot10^{-2}$ &$2.47\cdot10^{-3}$ &$1.92\cdot10^{-3}$ &$1.11\cdot10^{-4}$ &$8.63\cdot10^{-5}$  \\\hline
    6 \rule[-1pt]{0pt}{8pt}&$3.12\cdot10^{-1}$ &$3.92\cdot10^{-1}$ &$2.79\cdot10^{-2}$ &$3.50\cdot10^{-2}$ &$2.08\cdot10^{-3}$ &$2.62\cdot10^{-3}$ &$9.13\cdot10^{-5}$ &$1.15\cdot10^{-4}$  \\\hline
  \end{tabular}
  \caption{Cutflow for the signal processes. The cross section is given in picobarns after each cut. In the
  first line the parameter points are specified as $(m_{Q_3},m_{u_3})$, both in GeV.}
  \label{tab:cutflowSIG}
\end{table}

\begin{figure}[tb]
  \centering
  \subfloat[$\mu=150\,\GeV$]{\includegraphics[width=.47\textwidth]{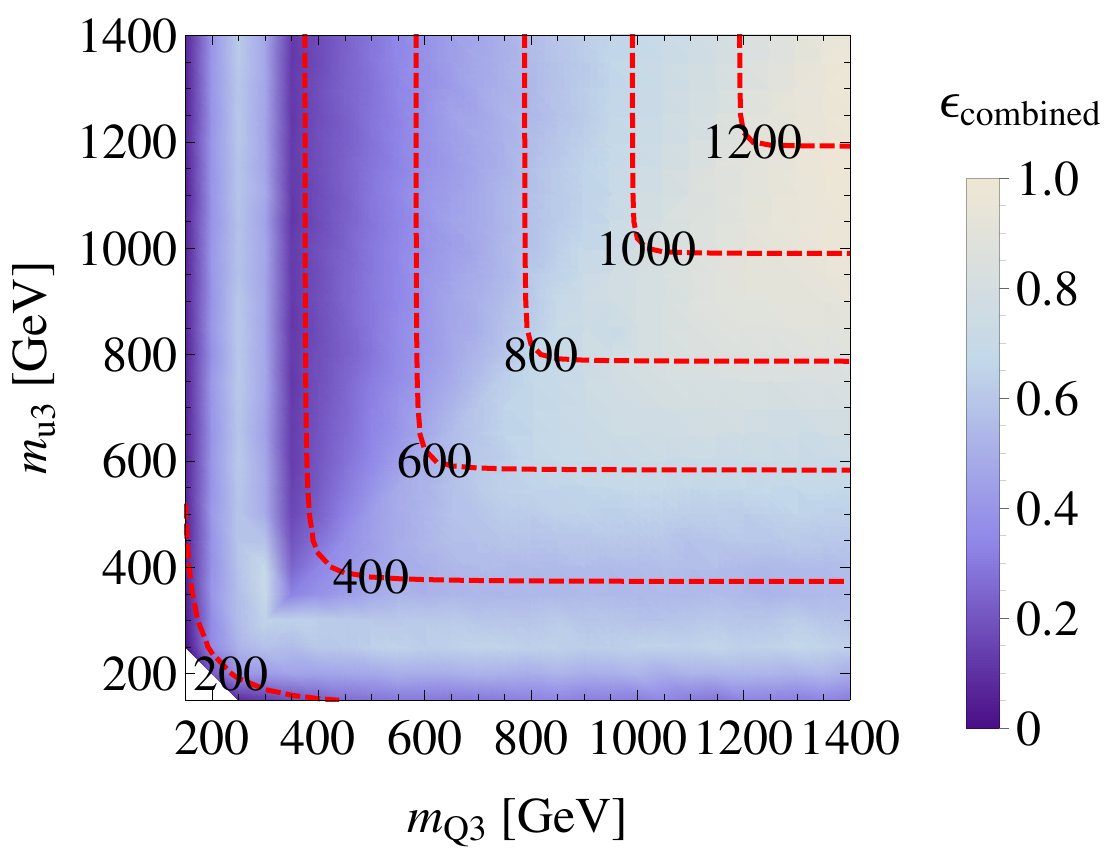}\label{fig:Efficinecy150}}
  \subfloat[$\mu=300\,\GeV$]{\includegraphics[width=.47\textwidth]{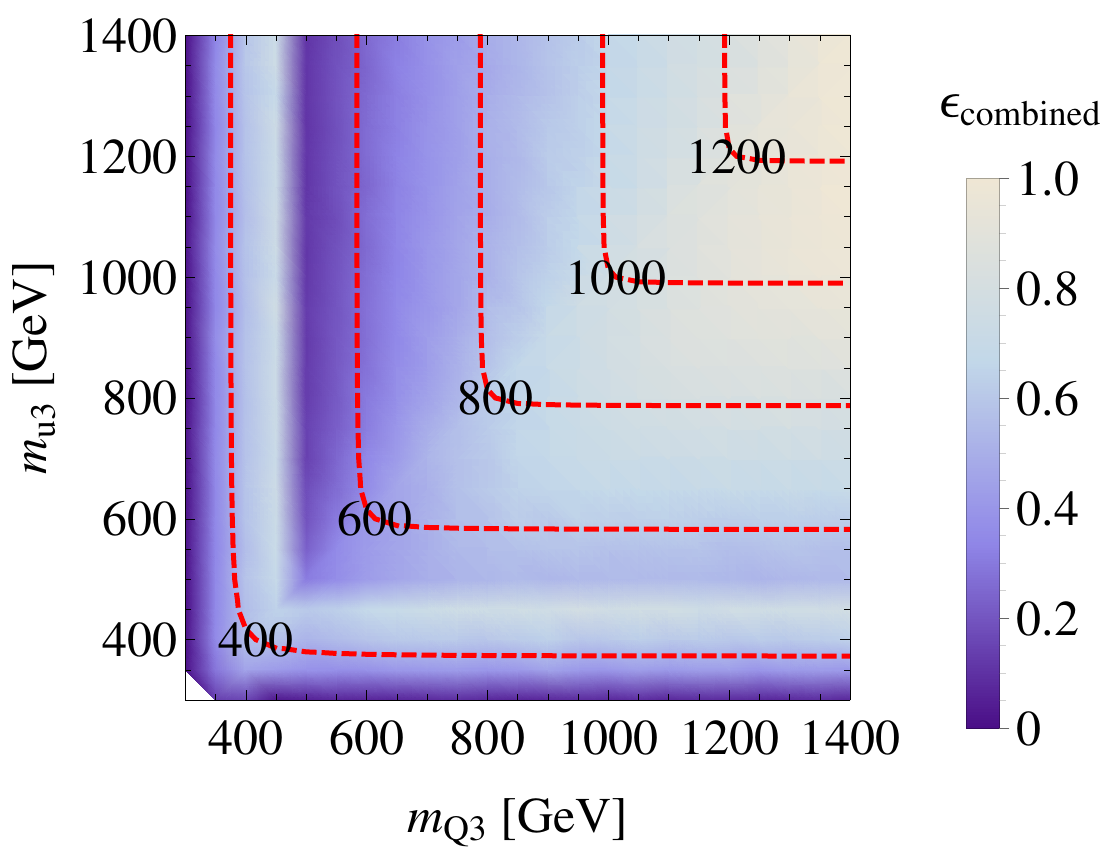}\label{fig:Efficinecy300}}
  \caption{Total efficiency of all cuts combined in the $m_{Q_3}$-$m_{u_3}$ plane. The red dashed
    lines show the mass of the lighter stop in GeV.}
  \label{fig:totalEfficiency}
\end{figure}

In the left plot in
Fig.~\ref{fig:NormedDistributions2} we show the relative contribution of the types of the two final
state candidates after all cuts. In the samples with heavy stops and sbottoms the HEPTopTagger
contributes between 30-40\% of the candidates. In the samples with lighter squarks and thus less
boosted objects the HEPTopTagger finds less candidates and the BDRS Tagger contributes up to about
15\% of the top quark candidates. However, most candidates besides the ones from the HEPTopTagger
are $b$-jets.\par
In Fig.~\ref{fig:Efficiency} we show the efficiency of each cut for the three possible final states
as a function of the squark mass. As anticipated they show only a mild mass dependence. This is also
reflected in the total efficiency which is very flat over the whole parameter space as can be seen
in Fig.~\ref{fig:totalEfficiency}.\par
The cutflows for background and signal together with the the signal over background ratios $S/B$ are
given in Tables \ref{tab:cutflowBKG} and \ref{tab:cutflowSIG}, respectively. The values for $S/B$
range from about 0.3 in the samples where the stops have a mass of only about $350\,\GeV$ to
$10^{-3}$ in the sample with heavy stops of about 1\,400\,GeV.\par

\subsection{Results}
\label{sec:discussion}

To continue further, we consider the $m_{T2}$ \cite{Lester:1999tx,Barr:2003rg} distribution that is
shown in the right plot of Fig.~\ref{fig:NormedDistributions2} (normalized) and in
Fig.~\ref{fig:StackedMT2} (stacked). $m_{T2}$ is designed to reconstruct the mass of the decaying
particle and gives a lower bound on it. This is reflected in the plotted distribution, where the
upper edge of the signal distribution is just at the actual squark mass.  For the calculation of
$m_{T2}$ we assume zero neutralino mass and use a code described in~\cite{Cheng:2008hk} and provided
by the authors of this reference.\par
Instead of imposing an explicit cut on $m_{T2}$ to improve $S/B$, mainly to the benefit of the
processes involving heavy squarks, we rather evaluate the statistical significance applying a binned
likelihood analysis using the $CL_s$ technique described in~\cite{Junk:1999kv,Read:2002hq}.  For the
calculation we employ the code \texttt{MCLimit} \cite{Junk:2007mcl}. We assume an uncertainty of
15\% on the background cross section and also include an error stemming from the finite size of the
Monte Carlo sample. For the latter we need to combine the pure statistical uncertainty with the
knowledge of a steeply falling background distribution. To do this we determine for each background
process and each bin the statistical uncertainty $\omega \sqrt{N}$, where $\omega$ is the weight of
one event and $N$ is the number of events in the given bin. Conservatively, we assign $N=1$ for
those bins which do not contain any events of the given process. In the high $m_{T2}$ region where
no background events appear this method clearly overestimates the error on the background which is
steeply falling. In addition we therefore fit the slopes of the $m_{T2}$ distributions with an
exponential function and use this function to extrapolate the background distribution to the
high-$m_{T2}$ region. As uncertainty on the shape for a given background process we now take in each
bin the minimum of $\omega \sqrt{N}$ and three times the fitted function. This way the error in the
low $m_{T2}$ range is determined by the statistical uncertainty while the one in the high $m_{T2}$
range from the extrapolation. The combined error on the background in each bin is then obtained by
summing the squared errors of each process and taking the square root.\par
The results are shown in Fig.~\ref{fig:CLsPlane150} (Fig.~\ref{fig:CLsPlane300}) for
$\mu=150\, (300) \,\GeV$ and integrated luminosities of $100, 300,$ and $1\,000\,\ifb$.  In
Fig.~\ref{fig:CLsvsLumi} we show for $(m_{Q_3},m_{u_3})=(1500,1500)$ the $CL_s$ exclusion limit as a
function of the integrated luminosity. Even for this parameter point, close to the predicted
sensitivity reach of the LHC, using our approach, we find a $95\%$ CL exclusion with
$600~\mathrm{fb}^{-1}$.

\begin{figure}[tb]
  \centering
    \includegraphics[height=128pt, trim=0pt 0pt 95pt 0pt, clip=true]{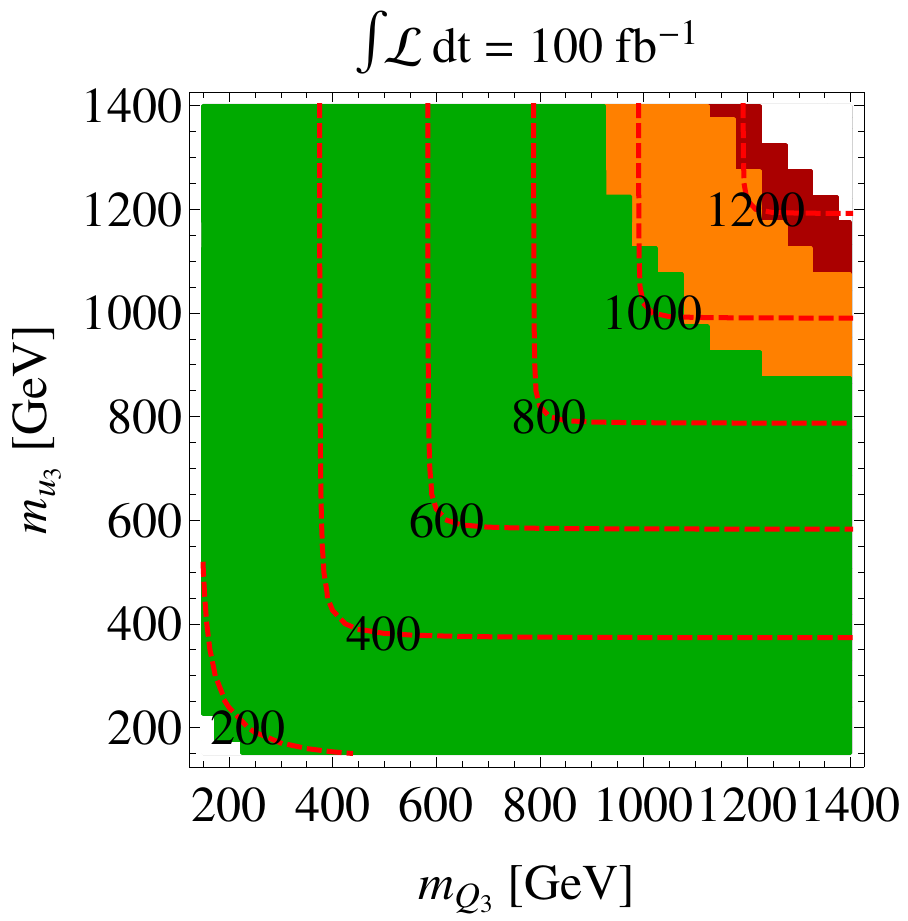}
    \includegraphics[height=128pt, trim=0pt 0pt 95pt 0pt, clip=true]{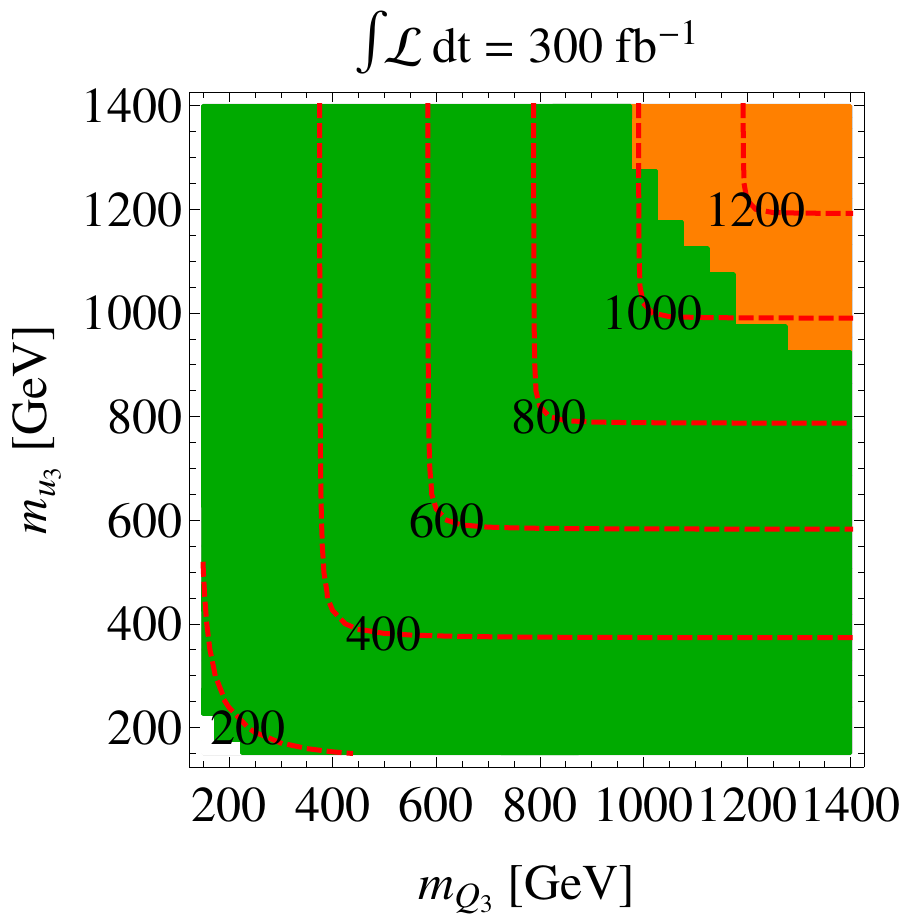}
    \includegraphics[height=128pt]{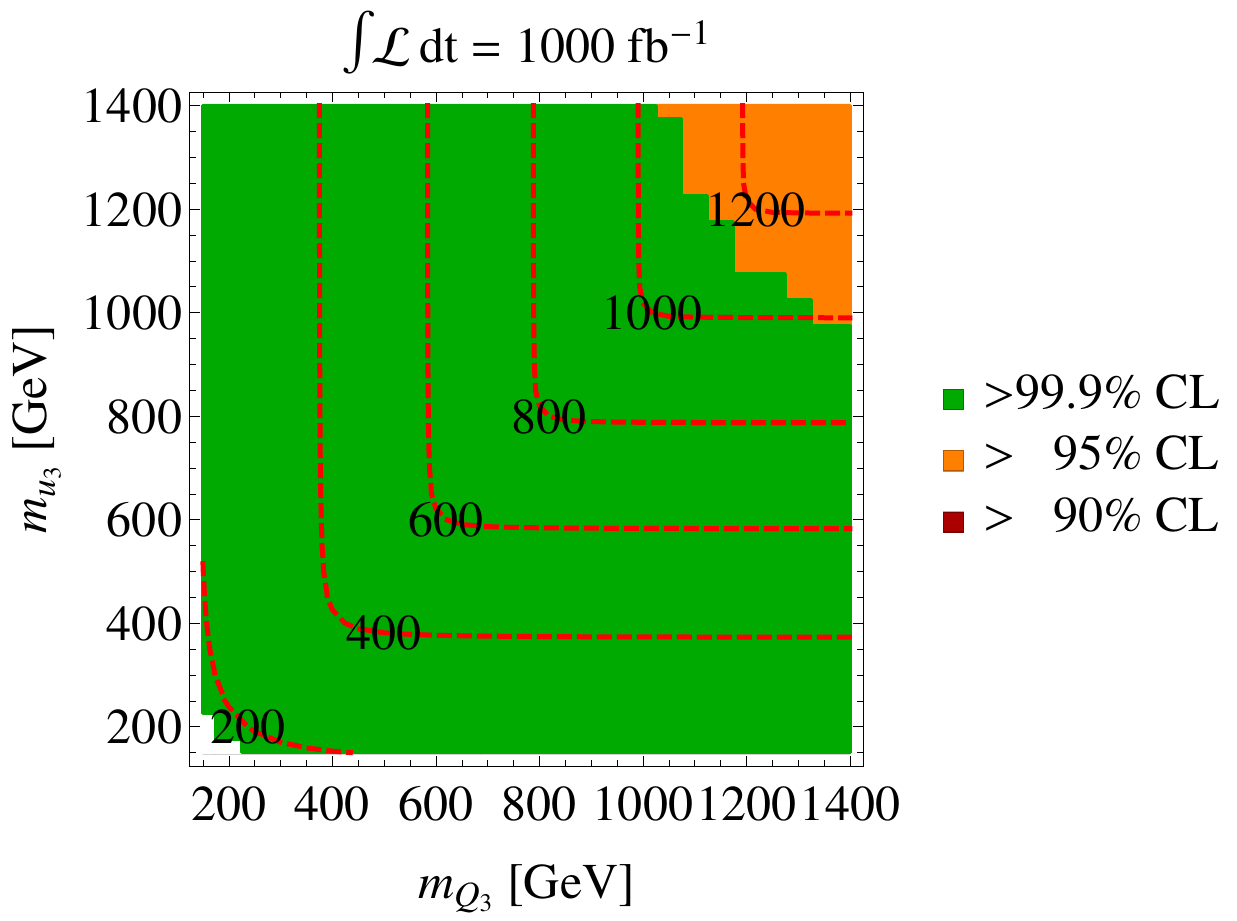}
    \caption{Exclusion limits in the $m_{Q_3}$-$m_{u_3}$-plane. A Monte Carlo error and a systematic
      error of 15\% on the background normalization is assumed. In all three plots $\mu=150\,\GeV$ which corresponds roughly to the mass of the higgsinos.}
  \label{fig:CLsPlane150}
\end{figure}
\begin{figure}[tb]
  \centering
    \includegraphics[height=128pt, trim=0pt 0pt 95pt 0pt, clip=true]{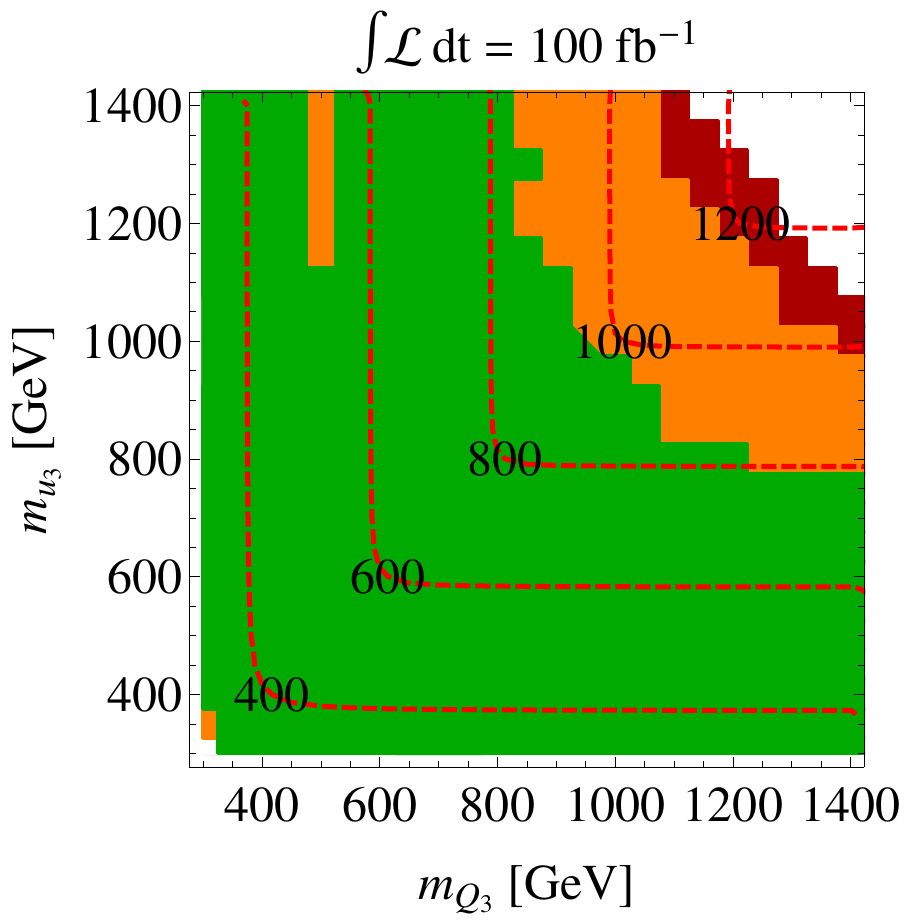}
    \includegraphics[height=128pt, trim=0pt 0pt 95pt 0pt, clip=true]{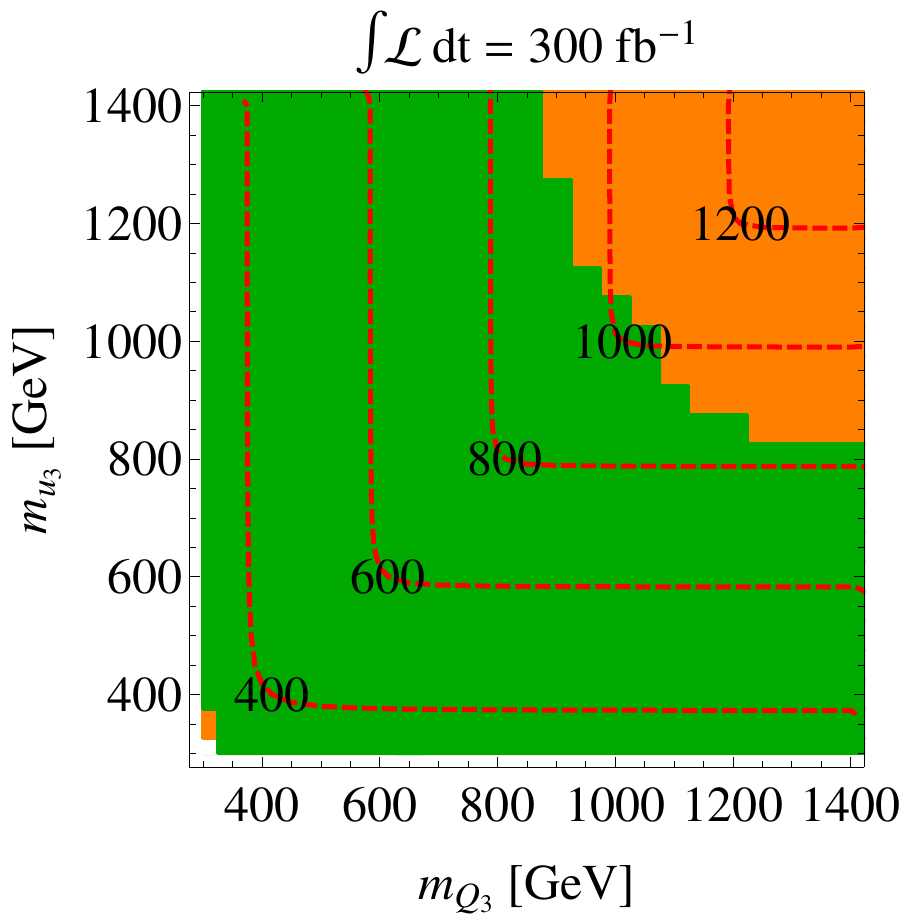}
    \includegraphics[height=128pt]{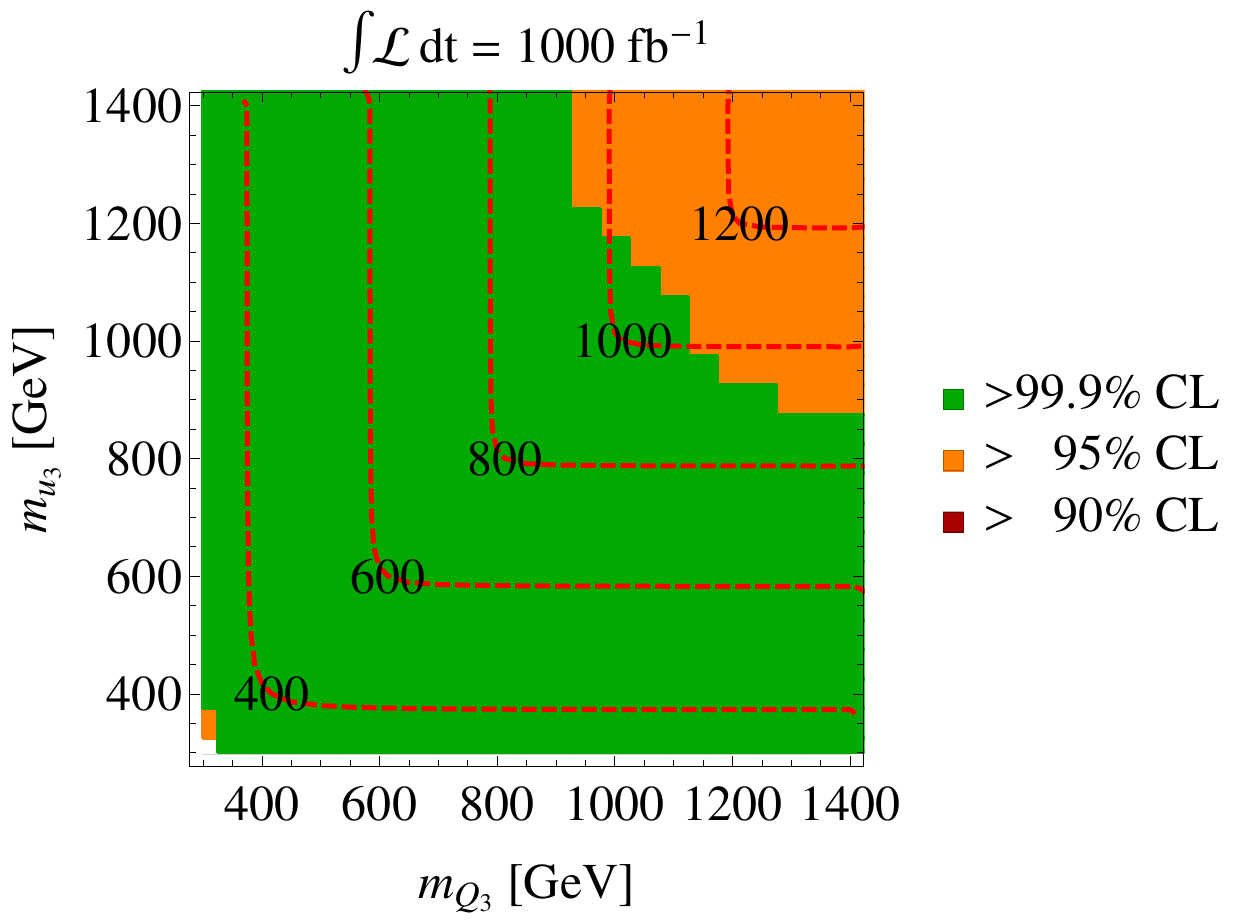}
    \caption{Exclusion limits in the $m_{Q_3}$-$m_{u_3}$-plane. A Monte Carlo error and a systematic
      error of 15\% on the background normalization is assumed as detailed in the main text. In all
      three plots $\mu=300\,\GeV$ which corresponds roughly to the mass of the higgsinos.}
  \label{fig:CLsPlane300}
\end{figure}

\begin{figure}[tb]
  \centering
  \includegraphics[width=0.5\textwidth]{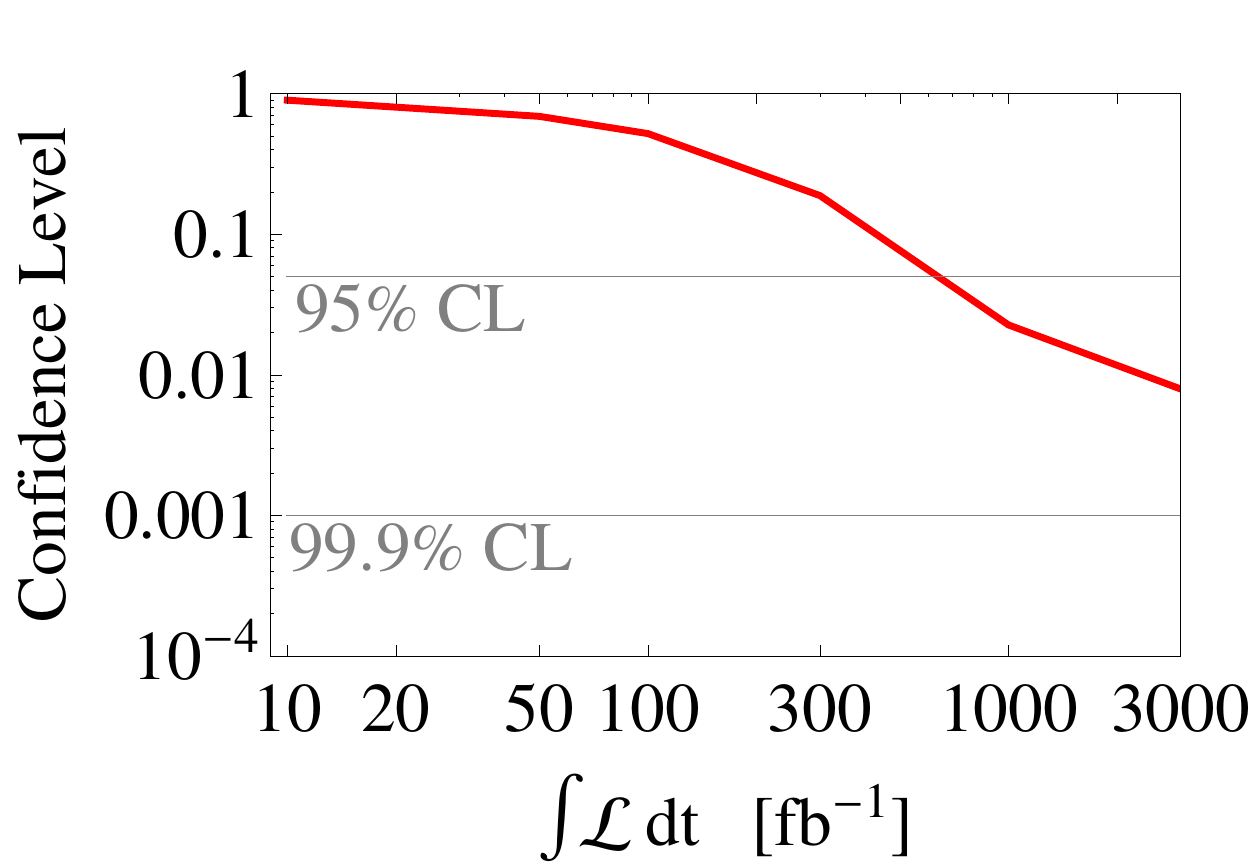}
  \caption{p-values as a function of integrated luminosity for the parameter point with
    $(m_{Q_3},m_{u_3})=(1500\,\GeV,1500\,\GeV)$ and $\mu=300\,\GeV$.}
  \label{fig:CLsvsLumi}
\end{figure}

\section{Final remarks}
\label{sec:final-remarks-1}

The main idea behind this analysis was to obtain a scale invariant setup. In the first step we
achieved this by employing the HEPTop and BDRS Taggers together with varying radii. Thereby we
managed to pick the minimal content of a hadronically decaying top quark for a large range of top
momenta. In the second step we avoided introducing scales in the cuts and only exploited the event
properties that are independent of the mass spectrum. After this proof of concept it will now be
interesting to apply this principle to other searches where top quarks with various boosts appear in
the final state as for example in little Higgs models with T-parity
\cite{Cheng:2003ju,Cheng:2004yc,Cheng:2005as}.

\section{Acknowledgments}

We are grateful to Rakhi Mahbubani and Minho Son for helpful discussions and support in the
initial phase of this project. AW and MSc thank the CERN TH division for hospitality and computing resources
and MSc thanks the Joachim Herz Stiftung for funding his work.

\clearpage{}
\addcontentsline{toc}{part}{Bibliography}
\bibliographystyle{utphys}
\bibliography{TheBib}
\markboth{}{}

\end{document}
